\documentclass[%
 reprint,
 superscriptaddress,
 amsmath,amssymb,
 aps,
 prb,
]{revtex4-2}
\usepackage{soul}
\usepackage{graphicx}
\usepackage{dcolumn}
\usepackage{bm}
\usepackage{xcolor}
\usepackage{multirow}
\usepackage{comment}
\usepackage{hyperref}

\begin{document}
\title{Classification of Dirac points with higher-order Fermi arcs}

\author{Yuan Fang}
\affiliation{Department of Physics and Astronomy, Stony Brook University, Stony Brook, New York 11974, USA}

\author{Jennifer Cano}
\affiliation{Department of Physics and Astronomy, Stony Brook University, Stony Brook, New York 11974, USA}
\affiliation{Center for Computational Quantum Physics, Flatiron Institute, New York, New York 10010, USA}
\date{\today}

\begin{abstract}
Dirac semimetals lack a simple bulk-boundary correspondence. 
Recently, Dirac materials with four-fold rotation symmetry have been shown to exhibit a higher order bulk-hinge correspondence: they display "higher order Fermi arcs", which are localized on hinges where two surfaces meet and connect the projections of the bulk Dirac points.
In this paper, we classify higher order Fermi arcs for Dirac semimetals protected by a rotation symmetry and the product of time-reversal and inversion.
Such Dirac points can be either linear in all directions or linear along the rotation axis and quadratic in other directions.
By computing the filling anomaly for momentum-space planes on either side of the Dirac point, we find that all linear Dirac points exhibit higher order Fermi arcs terminating at the projection of the Dirac point, while the Dirac points that are quadratic in two directions lack such higher order Fermi arcs.
When higher order Fermi arcs do exist, they obey either a $\mathbb{Z}_2$ (four-fold rotation axis) or $\mathbb{Z}_3$ (three- or six-fold rotation axis) group structure.
Finally, we build two models with six-fold symmetry to illustrate the cases with and without higher order Fermi arcs. We predict higher order Fermi arcs in Na$_3$Bi.
\end{abstract}

\maketitle


\section{\label{sec:intro}Introduction}

{Topological semimetals encompass a large family of materials exhibiting band crossings near the Fermi level \cite{wieder2021topological}, such as Weyl \cite{wan2011topological,weng2015weyl,huang2015weyl,xu2015discovery,lv2015observation,xu2015discovery2,lv2015experimental,xiong2015evidence}, Dirac \cite{young2012dirac,wang2012dirac,liu2014discovery,liu2014stable,steinberg2014bulk} and multifold fermions \cite{bradlyn2016beyond,wieder2016double,cano2019multifold}.}
One of the novel features of Weyl and other chiral semimetals is their bulk-edge correspondence in the form of surface Fermi arcs \cite{wan2011topological}.
The surface Fermi arcs are a direct consequence of the nontrivial bulk topology of a Weyl fermion, i.e., that it is a source of Berry curvature.

\begin{figure*}
    \centering
    \includegraphics[width=0.9\linewidth]{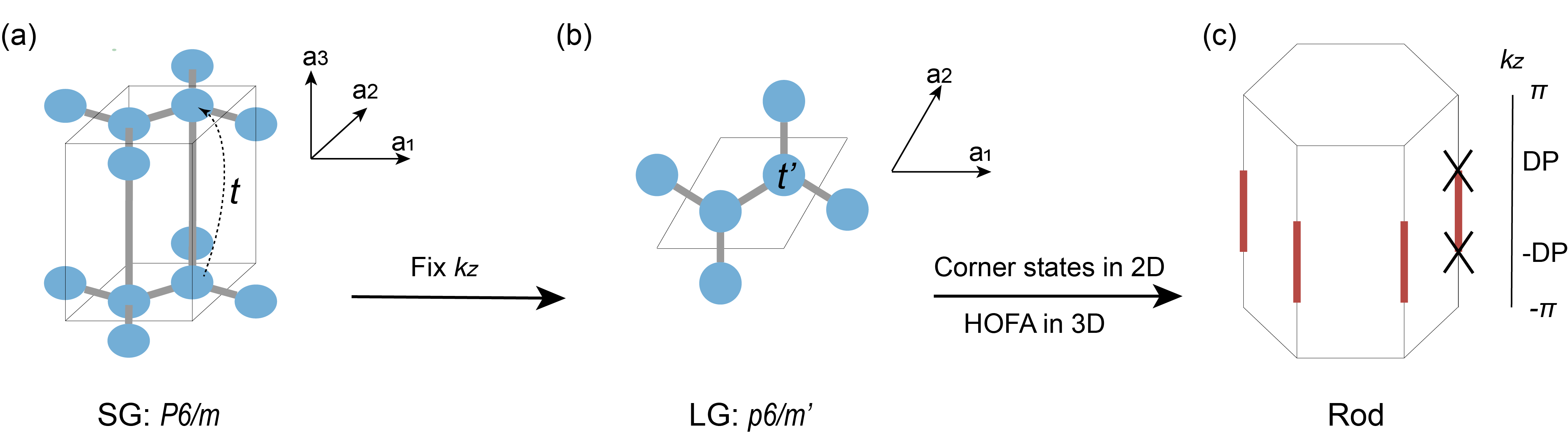}
    \caption{{Schematic diagram showing HOFAs.}
    (a) Hexagonal unit cell of space group $P6/m$. 
    (b) Each plane with fixed $k_z\neq 0,\pi$ is regarded as an effective 2D system with the symmetry of the magnetic layer group $p6/m'$. 
    After this dimensional reduction, interlayer hopping between atoms separated by a unit cell is viewed as a {$k_z$-dependent} onsite potential: $t'=te^{ik_z}+t^\dagger e^{-ik_z}$. 
    Other interlayer hopping terms are similarly projected to in-plane hopping terms.
    (c) When the crystal is terminated in a $C_6$-symmetric rod geometry, the 2D planes with nontrivial filling anomaly contribute corner states to HOFAs of the 3D model.
    Red lines indicate HOFAs that could appear between two Dirac points (whose projection onto the hinges is labelled by crosses).}
    \label{fig_scheme}
\end{figure*}

A similar bulk-edge correspondence does not exist for Dirac semimetals because they are not a source of Berry curvature \cite{kargarian2016surface,kargarian2018deformation,le2018dirac}.
However, recently, certain Dirac semimetals have been shown to have a higher order bulk-edge correspondence, in the form of higher order Fermi arcs (HOFAs) \cite{wieder2020strong}.
HOFAs are 1D mid-gap modes that are localized on the ``hinges'' of a crystal where two surfaces meet \cite{wieder2020strong,lin2018topological,ghorashi2020higher}.
The HOFAs connect the projection of the bulk Dirac points in the 1D rod BZ (i.e., the BZ of a crystal finite in two dimensions and infinite in the third),
analogous to how surface Fermi arcs connect the projection of bulk Weyl points on the surface BZ.
An example is shown in Figure~\ref{fig_scheme}.
Ref.~\cite{wieder2020strong} proved that for Dirac fermions in a crystal with a four-fold rotation symmetry, HOFAs are required.
Thus, the HOFAs furnish a ``bulk-hinge correspondence'' for these Dirac semimetals.

Refs~\cite{wieder2020strong,lin2018topological} were limited to the case of four-fold rotational symmetry and relied on the nontrivial quadrupole index \cite{benalcazar2017electric,benalcazar2017quantized} specific to that case. However, Dirac cones can also be protected by three- and six-fold rotations.
In this work, we derive a more general formulation of the bulk-hinge correspondence that applies to any symmetry group, thus 
answering the question: when do Dirac fermions have HOFAs?

Our strategy is to compute the filling anomaly~\cite{benalcazar2019quantization,schindler2019fractional} for 2D symmetry-preserving momentum-space slices.
A nontrivial filling anomaly in an insulating plane requires the existence of mid-gap states localized at corners where two edges meet.
Thus, when the filling anomalies of planes on either side of the Dirac point are different, the plane on at least one side of the Dirac point must have a non-zero filling anomaly and accompanying corner states.
The corner states coming from adjacent planes together form the HOFA.
Thus, the bulk-hinge correspondence that results from the change in filling anomaly across the Dirac point is analogous to the bulk-edge correspondence that results from the change in Chern number across a Weyl point.

Our main result is that all linear Dirac points protected by a three-, four- or six-fold rotation symmetry, along with the product of time-reversal and inversion symmetries, have HOFAs that terminate at the projection of the Dirac point; we use linear Dirac point to refer to a dispersion that is linear in all three directions.
A six-fold rotation symmetry can also protect quadratic Dirac points, which are linear along the rotation axis but quadratic in other directions; we find that the quadratic Dirac points do not have HOFAs.
These results are derived in Sec.~\ref{sec:Filling}.
An additional consequence of our work is that multifold fermions at time-reversal invariant momenta \cite{bradlyn2016beyond} do not have HOFA that terminate at the projection of the bulk gapless point because such HOFA would violate time-reversal symmetry. Thus, it remains to find a bulk-edge or bulk-hinge correspondence for these fermions. 

Our results serve as a topological classification: a Dirac fermion can be classified by how the filling anomaly changes across the Dirac point, which is summarized in Table~\ref{tab:classification}.
We now compare this classification to previous classifications of Dirac points.
In Ref.~\cite{yang2015topological}, Dirac points are classified by the change in symmetry of the valence bands adjacent to the Dirac point.
This is similar to our approach, in that we both examine the valence bands adjacent to the Dirac point.
However, the classification in Ref.~\cite{yang2015topological} is a classification in momentum space, while our classification uses momentum space eigenvalues to determine the position of Wannier centers in real space.
As a result, the classifications differ in their definition of the trivial phase: in Ref.~\cite{yang2015topological}, the trivial phase corresponds to the absence of a Dirac point, 
while the trivial phase in our classification is defined by the filling anomaly remaining constant across a particular plane and, equivalently, no HOFAs that terminate at the projection of the Dirac point.

In Ref.~\cite{yang2014classification}, Dirac points are classified by the change in the 2D topological invariant of each TRIM plane (i.e., the $\mathbb{Z}_2$ and mirror Chern indices) when a Dirac point is between them.
We rederive and extend these results using topological quantum chemistry \cite{bradlyn2017topological,vergniory2017graph,elcoro2017double,bradlyn2018band,cano2018building,vergniory2019complete} (Sec.~\ref{sec:TRIM}) and compare this classification to ours in Table~\ref{tab_summary} (Sec.~\ref{sec:connect}).
Recently, a third classification was introduced in Ref.~\cite{tyner2020topology}, where Dirac points are classified by their non-Abelian Berry flux.
It remains an open question to determine whether the physical observables of this classification correspond to higher order Fermi arcs.

Finally, we illustrate our results with several examples.
In Sec.~\ref{sec:example}, we build two four band models protected by six-fold rotation symmetry, with and without HOFAs.
We discuss the application to Na$_3$Bi and predict the existence of higher order Fermi arcs.

\par

\section{\label{sec:Filling} Higher order Fermi arcs via the filling anomaly}

As explained in Sec.~\ref{sec:intro}, three dimensional Dirac semimetals do not have topologically protected surface Fermi arcs~\cite{kargarian2016surface,kargarian2018deformation}, but can have 1D HOFAs on hinges, as illustrated schematically in Figure~\ref{fig_scheme}(c).

At the crux of our analysis of HOFAs is the filling anomaly~\cite{benalcazar2019quantization,schindler2019fractional}. The filling anomaly $\eta$ is defined for an insulator on a symmetrically terminated lattice that is in an atomic limit phase (i.e. admits localized symmetric Wannier functions \cite{bradlyn2017topological}). 
It is given by the difference between the ion charge and the electron charge in the valence bands:
\begin{equation}
    \eta = |\#\text{ion}|-|\#\text{electron}|,
    \label{eq:defeta}
\end{equation}
in units of the electron charge $|e|$.
The filling anomaly is only defined modulo an integer given by the minimal number of electrons that can be added or removed to the boundary of the system while preserving crystal symmetry.
For example, in a finite 2D lattice with four-fold rotational symmetry, one can always add (or remove) four electrons to the corners; thus, the filling anomaly is defined mod 4. In the presence of time-reversal symmetry, electrons must be added in pairs, and the filling anomaly would be defined mod 8.
The filling anomaly also depends on the choice of lattice termination~\cite{fang2021filling}.
In the presence of an $n$-fold rotation symmetry, we will always consider a lattice termination whose cross-section is a regular $n$-gon.

When $\eta$ in Eq.~(\ref{eq:defeta}) is nonzero, it means that the finite-size insulator cannot be both neutral and symmetric.
This conflict is resolved by the presence of mid-gap states.
{If the crystal is bulk-insulating and has no polarization (surface charge), then the mid-gap states must be localized at corners.}
Thus, the filling anomaly defines the corner charge \cite{benalcazar2019quantization}:
\begin{equation}
    Q_c=\frac{\eta}{n}|e|
    \label{eq:defQc}
\end{equation}
for a crystal with an $n$-fold rotation axis.
Since we do not impose particle-hole {or chiral} symmetry, the corner states are not required to be at exactly zero energy; further, their energy can be manipulated by a surface or corner potential.
Nonetheless, the filling anomaly is robust because it is defined modulo the number of corner states and thus does not change when energy of the corner states changes \cite{wieder2020strong}.

The filling anomaly in 2D can be a useful tool to understand higher order topological insulators \cite{benalcazar2017quantized,benalcazar2017electric,song2017d,Khalaf2018symmetry,Khalaf2018higher,schindler2018higher,schindler2018higherBi,Song2017,Langbehn2017,Geier2018,Imhof2017,Peterson2018,SerraGarcia2018,Noh2018,Trifunovic2019higher,fang2020higher,fang2021filling}, as well as semimetals, in 3D.
For example, a $\mathbb{Z}_8$ higher order topological insulator has a filling anomaly $\eta=4\mod 8$ at one of its two TRIM planes ($k_z=0,~\pi$), indicating in which plane the helical hinge modes cross \cite{fang2021filling,song2017d}.
As discussed in Sec.~\ref{sec:intro}, the filling anomaly also explains HOFAs.
Specifically, in a 3D semimetal, the 2D momentum-space planes with fixed $k_z$ that do not contain the Dirac point are 2D insulators.
If the filling anomaly of a 2D plane is non-zero, then it will have mid-gap corner states.
Since the filling anomaly is robust to perturbations that do not close the gap, the filling anomaly must remain constant when continuously varying $k_z$.
The corner states from all such planes make up the HOFA.
In four-fold symmetric Dirac semimetals, it was shown that the filling anomaly changes when $k_z$ moves across the Dirac point \cite{wieder2020strong,fang2021filling}.
Thus, the HOFAs are terminated by the planes containing the Dirac points.

In this work, we will generalize this logic to determine when Dirac points have HOFAs in other space groups.
Specifically, HOFAs begin/end on a Dirac point exactly when the filling anomaly of 2D momentum-space planes changes across it. 
We will classify the Dirac points that have this property. By doing so, we show that not all Dirac points have HOFAs. 
We remark here that such a jump is only a sufficient condition. If there is a non-zero filling anomaly on both sides of the Dirac point, then there will also be HOFAs; however, these HOFAs are not associated with the Dirac points and do not terminate at them.

\subsection{\label{sec:m_layer_group} Symmetry of $k_z$ slices: magnetic layer groups}

We consider Dirac points protected by the combination of time reversal, $\mathcal{T}$, with $\mathcal{T}^2 = -1$, inversion, $\mathcal{I}$, and an $n$-fold rotation symmetry, $C_{n}$, with $n=3,4,6$.
We discuss these and other symmetries that can protect Dirac points in more detail in Appendix~\ref{app:DP}.
We are only interested in Dirac points not at a TRIM, since the filling anomaly, $\eta$, will not change across these Dirac points because $\eta$ is invariant under time-reversal.
Notice this implies HOFAs cannot terminate  at the projections of multifold fermions at TRIMs in time-reversal symmetric systems. This does not contradict Ref.~\cite{nie2021sixfold} where the HOFAs do not terminate  at the six-fold fermion.

We take the $n$-fold rotation axis to be the $\hat{z}$ direction. Thus, to determine the presence/absence of HOFA, we will compute the filling anomaly for 2D planes in the BZ with fixed $k_z$.
The symmetry group of a 2D slice of the BZ is described by a layer group.
Generic slices ($k_z\neq 0,~\pi$) are invariant under the product $\cal TI$ and $C_{n=3,4,6}$.
Since these slices are not time-reversal invariant, they are described by the magnetic layer groups, $p\bar{3}'$, $p4/m'$, and $p6/m'$, respectively.
If the crystal has a mirror symmetry (in addition to $\cal T$, $\cal I$, and $C_n$), the same analysis of the filling anomaly applies (see Appendices~\ref{app:DP_Cn} and~\ref{app:wyck}).
The corresponding magnetic layer groups are: $p4/m'mm$, $p\bar{3}'m1$, $p\bar{3}'1m$, and $p6/m'mm$.

Our analysis also applies to Dirac cones protected by an $n$-fold screw symmetry along the $z$-axis instead of an $n$-fold rotation symmetry
because the little co-groups along the high symmetry lines in the case of screw symmetry are isomorphic to the little co-groups in the case of rotation symmetry, and therefore the irreducible representations (irreps) are identical (up to a phase that results from the translation), as long as $k_z\neq \pi$ \cite{bradley2009mathematical}. 
We discuss this point in more detail in Appendix~\ref{app:DP_snp}.

However, our analysis does not apply to non-symmorphic groups that protect a Dirac point at the boundary of the BZ \cite{young2012dirac,wieder2016double,bradlyn2016beyond,cano2019multifold}.
We discuss examples in Appendix~\ref{sec_DP_discuss}.

\subsection{\label{sec:Filling_filling}Filling anomaly of $k_z$ slices}

The filling anomaly is defined for finite systems. 
{However, it is more efficient to compute the filling anomaly from bulk invariants, which is accomplished via the bulk-corner correspondence.}
Specifically, the filling anomaly can be determined by the number of bulk Wannier functions 
centered at each maximal Wyckoff position. 
Analogous to the well known relations between the polarization, Wannier centers and Berry phase in 1D \cite{resta1994macroscopic,vanderbilt1993electric,vanderbilt2018berry}, 
recently, the relation between the filling anomaly and Wannier centers in 2D has been derived in generality~\cite{watanabe2020corner,fang2021filling,takahashi2021general}. It can be expressed succinctly as
\begin{align}
\label{eqn_eta_nw}
    \eta = a_a-e_a~ \mod r ~(\text{or}~ 2r),
\end{align}
where $a_a$ ($e_a$) is the number of atoms ({electron Wannier centers}) at the Wyckoff position $1a$. The Wyckoff positions of the relevant magnetic layer groups (discussed in Sec.~\ref{sec:m_layer_group}) are reviewed in Appendix~\ref{app:wyck}. 
\par

The modulus $r(2r)$ in Eq.~(\ref{eqn_eta_nw}) enters because the filling anomaly is defined modulo the minimum number of electrons that can be added to the finite system while preserving symmetry (as explained below the definition of $\eta$ in Eq.~(\ref{eq:defeta})).
The choice of $r$ in Eq.~(\ref{eqn_eta_nw}) applies to a 2D system invariant under the product $\cal TI$, but not under $\cal T$ and $\cal I$ separately; the choice of $2r$ in in Eq.~(\ref{eqn_eta_nw}) applies when both $\cal T$ and $\cal I$ are symmetries.
The value of $r$ depends on the index of the rotation: if the crystal has four-fold rotation symmetry, $r=4$, while if it has three- or six-fold rotation symmetry, $r=6$.
In Appendix~\ref{app:r} we derive the modulus $r(2r)$ using group theory.

When zero-dimensional mid-gap states exist, their degeneracy must be equal to the modulus of the filling anomaly ($r$ or $2r$) in order to preserve crystal symmetry.
(The value of $\eta$ determines how many electrons are available to fill these $r$ or $2r$ states.)
In the presence of $C_n$ ($n=3,4,6$), $\cal T$, and $\cal I$ symmetries (i.e., at a TRIM plane), the degeneracy of the corner states jumps from $r$ to $2r$.
The additional states could come from gapless surface states (if the TRIM plane has a nontrivial $\mathbb{Z}_2$ strong 2D TI index or mirror Chern number, as we discuss in Sec.~\ref{sec:TRIM}), bulk states, or from another set of hinge states.

{We now explain how to obtain the atomic positions and electron Wannier centers that enter Eq.~(\ref{eqn_eta_nw}), which will be applied to each $k_z$-slice.
To obtain the atomic positions, each atom in the unit cell is projected to the $z=0$ plane.
Interlayer hopping terms then become $k_z$-dependent in-plane hopping terms in the 2D (fixed $k_z$) Hamiltonian.}
In Figure~\ref{fig_scheme} we show an example of this process in space group $P6/m$.
In this example, the unit cell compatible with the rod geometry is a hexagonal unit cell, where the $a_3$ lattice vector is in the $z$-direction and is perpendicular to the other two lattice vectors.
The effective 2D model at fixed $k_z\neq 0,\pi$ (see Figure~\ref{fig_scheme}(b)) has an effective 2D lattice with layer group $p6/m'$. The $k_z$-dependent onsite term $t'$ of the 2D model is determined by the inter-plane hopping $t$ of the 3D model.
The localization of the electrons in the $a_1-a_2$ plane 
can be determined by this 2D model. According to Eq.~(\ref{eqn_eta_nw}), the localization center determines the filling anomaly, which determines the presence/absence of HOFAs via the corner charge formula in Eq.~(\ref{eq:defQc}).

{
If the 3D crystal has different planes with atoms in different positions, all atoms in all planes in the unit cell should be projected to $z=0$ to obtain the Wyckoff positions that enter Eq.~(\ref{eqn_eta_nw}). Thus, the atomic positions will be the same for each $k_z$-slice (even though the atomic positions in different real space planes with fixed $z$ may differ).}
An example with multiple distinct planes in the unit cell is shown in Figure~\ref{fig_scheme2} for space group $P6_3/m$ in Appendix~\ref{app:DP}.\par

\subsection{\label{sec:Filling_indicator} Symmetry indicators for filling anomalies}

The Wannier centers that enter Eq.~(\ref{eqn_eta_nw}) can be determined from the Wannier functions.
However, this process is computationally intensive and can often be simplified by symmetry indicator formulas, where
the filling anomaly is expressed in terms of the number of times each irreducible co-representation (co-irrep) appears at each high symmetry point. 

The symmetry indicator formula for the filling anomaly has been derived for many 2D symmetry groups~\cite{benalcazar2019quantization,schindler2019fractional,takahashi2021general,fang2021filling}.
In Ref.~\cite{fang2021filling}, we derived an algorithm to generate the symmetry indicators by building on the theory of topological quantum chemistry \cite{bradlyn2017topological}.
There, we found that the number of electrons whose Wannier centers are at each Wyckoff position is determined by the Smith normal form {of the so-called EBR matrix} (see Appendix~\ref{app_EBR} for details).
This approach determines the number of Wannier centers at each Wyckoff position {only up to some modulus,} which indicates that the mapping from symmetry co-irreps to Wannier centers is not one to one. 
{We will use our algorithm to compute the symmetry indicator formulas for the magnetic layer groups of interest.} \par

As discussed in Sec.~\ref{sec:m_layer_group}, we are interested in three layer groups: $p4/m'$, $p\bar{3}'$, and $p6/m'$. 
We compute the symmetry indicator formulas for each Wyckoff position in these groups in Appendix~\ref{app_EBR}.
The co-irreps are labelled by the subscript $j_z$, which determines the $C_n$ eigenvalues of a particular co-irrep by $\xi = e^{\pm i2\pi j_z/n}$ (generalizing the notation of Altmann and Herzig~\cite{altmann1994point} to the co-irreps of magnetic point groups).
We now summarize the results:
\par

\begin{table}[]
\begin{tabular}{c|cc}
$4/m'$ &$E$&$C_4$\\
\hline
$E_{\frac12}$&$2$&$\sqrt{2}$\\
$E_{\frac32}$&$2$&$-\sqrt{2}$
\end{tabular}
\quad
\begin{tabular}{c|ccc}
$6/m'$ &$E$&$C_6$&$C_3$\\
\hline
$E_{\frac12}$&$2$&$\sqrt{3}$&$1$\\
$E_{\frac32}$&$2$&$0$& $-2$\\
$E_{\frac52}$&$2$&$-\sqrt{3}$&$1$
\end{tabular}
\quad
\begin{tabular}{c|cc}
$\bar{3}'$ &$E$&$C_3$\\
\hline
$E_{\frac12}$&$2$&$1$\\
$E_{\frac32}$&$2$&$-2$
\end{tabular}
\caption{Character tables of the spinful irreducible co-representations of magnetic point groups $4/m'$, $6/m'$ and $\bar{3}'$.}
\label{tab_PT_irrep}
\end{table}

\begin{description}
    \item[$p4/m'$] The symmetry indicators giving the number of electrons with Wannier centers at Wyckoff positions $1a$ and $1b$ are:
\begin{align}
\label{eqn_na1}
    e_a &= N-2[M_{\frac12 }]&\mod 4,\qquad\\
\label{eqn_nb1}
    e_b &= 2[M_{\frac12 }]&\mod 4,\qquad
\end{align}
where N is the number of filled bands 
and $[M_{\frac12}]$ is the difference in the number of times the co-irrep $E_\frac12$ appears at $M=(\pi,\pi)$ and at $\Gamma=(0,0)$ in the valence bands. (The {little co-group} of both $\Gamma$ and $M$ is $4/m'$; its co-irreps are listed in Table~\ref{tab_PT_irrep}.)
We find $e_c=0$ mod 4.
\par
Plugging Eq.~(\ref{eqn_na1}) into Eq.~(\ref{eqn_eta_nw}) yields the symmetry indicator formula for the filling anomaly:
\begin{align}
\label{eqn_eta1}
    \eta^{(4)} = a_a-N+2[M_{\frac12}]\mod 4,
\end{align}
where the superscript 4 indicates the four-fold rotation symmetry.

\item[$p6/m'$] The symmetry indicators giving the number of electrons with Wannier centers at Wyckoff positions $1a$ and $2b$ are:
\begin{align}
\label{eqn_na2}
    e_{a}&=N - 2[K_{\frac12}] &\mod 6, \\
\label{eqn_nb2}
    e_{b}&=2[K_{\frac12}] &\mod 6,
\end{align}
where N is the total number of filled bands and
$[K_{\frac12}] = \# K_\frac12-\# \Gamma_\frac12 - \# \Gamma_\frac52$, where $\# P_\rho$ indicates the number of times the irrep $\rho$ appears at the high-symmetry point $P$ in the valence bands.
The little co-group at $K=(4\pi/3,0)$ is $\bar 3'$ and the little co-group at $\Gamma=(0,0)$ is $6/m'$. 
Their co-irreps are listed in Table~\ref{tab_PT_irrep}.\par
Plugging Eq.~(\ref{eqn_na2}) into Eq.~(\ref{eqn_eta_nw}) yields the symmetry indicator formula for the filling anomaly:
\begin{align}
\label{eqn_eta2}
    \eta^{(6)} = a_a-N+2[K_{\frac12}] \mod 6.
\end{align}
where the superscript 6 indicates the six-fold rotation symmetry.

\item[$p\bar{3}'$] 
The symmetry indicators giving the number of electrons with Wannier centers at Wyckoff positions $1a$ and $2b$ are
\begin{align}
    \label{eqn_na3}
    e_a &= N + 2[K_{\frac12}] + 2[K'_{\frac12}] &\mod 6,\\
    \label{eqn_nb3}
    e_b &= -2[K_{\frac12}] - 2[K'_{\frac12}] &\mod 6,
\end{align}

where N is the total number of filled bands,
$[K_{\frac12}] = \# K_\frac12-\# \Gamma_\frac12$, $[K'_{\frac12}] = \# K'_\frac12-\# \Gamma_\frac12$ and $\# P_\rho$ indicates the number of times the irrep $\rho$ appears at the high-symmetry point $P$ in the valence bands.
The little co-groups at $\Gamma=(0,0)$, $K=(4\pi/3,0)$ and $K'=(-4\pi/3,0)$ are all $\bar{3}'$. 
Their co-irreps are listed in Table~\ref{tab_PT_irrep}.

Plugging Eq.~(\ref{eqn_na3}) into Eq.~(\ref{eqn_eta_nw}) yields the symmetry indicator formula for the filling anomaly:
\begin{equation}
\label{eqn_eta3}
    \eta^{(3)} = a_a-N-2[K_{\frac12}]-2[K'_{\frac12}] \mod 6,
\end{equation}
where the superscript 3 indicates the three-fold rotation symmetry.

\end{description}

\subsection{\label{sec:Filling_class} Classification of Dirac points}
 

We classify each Dirac point by how the filling anomaly of a 2D fixed-$k_z$ slice changes across the Dirac point.
Specifically, if the Dirac point is at some $k_z = k_0$, then nearby planes at $k_z^{(-)}=k_0-\delta k_z$ and $k_z^{(+)}=k_0+\delta k_z$ are insulating for small $\delta k_z$. 
A 2D insulating system with only $C_n$ and $\cal TI$ symmetries has no symmetry indicated stable topological phase, as we compute in Appendix~\ref{app_EBR} by applying topological quantum chemistry to the magnetic layer groups.
As a result, the 2D systems in the planes $k_x^{(\pm)}$ are either in an atomic limit or fragile \cite{po2018fragile,cano2018topology} phase.
{In an atomic limit phase, we could compute the exponentially localized Wannier functions and determine the number of Wannier centers at each Wyckoff position. 
Below, we will describe how to compute the filling anomaly from this data.
Although the fragile phases lack exponentially localized Wannier functions, we can deduce the filling anomaly for the fragile phases from that of the atomic limit phases because each fragile phase can be expressed as a ``subtraction'' between two atomic limit phases.}

From the number of Wannier centers $n_w$ at each Wyckoff position, we can determine the filling anomaly for each plane, from which we define the
change in filling anomaly:
\begin{equation}
    \Delta \eta = \eta(k_z^{(+)})-\eta(k_z^{(-)}).
    \label{eq:defdeltaeta}
\end{equation}
We can express $\Delta \eta$ using symmetry indicators by applying Eqs.~(\ref{eqn_eta1}), (\ref{eqn_eta2}) and (\ref{eqn_eta3}). The results, which we will derive shortly, are summarized in the third and fourth columns in Table~\ref{tab:classification}.

The classifications for $n=4,6$ and $3$ are $\mathbb{Z}_2$, $\mathbb{Z}_3$ and $\mathbb{Z}_3$, respectively. 
We interpret this group structure by assigning each Dirac point a group element, or ``charge'', based on the change in filling anomaly between planes adjacent to the Dirac point (Eq.~(\ref{eq:defdeltaeta})).
Two Dirac points can add in the sense that if they are both within the range $(k_z^{(+)},k_z^{(-)})$, then the change in filling anomaly defined by Eq.~(\ref{eq:defdeltaeta}) will be the sum of the charges of each Dirac point.
If tuning some parameter in the Hamiltonian creates a pair of crossings between the conduction and valence bands, the Dirac points at the crossings will have opposite co-irreps in their valence bands and thus
opposite $\Delta \eta$; together their charges add to zero and these Dirac points are inverses of each other.
However, it may be that the combined charge of two Dirac points is zero even if they do not annihilate each other. Then, these Dirac points are inverses according to the group structure, but if they are fine-tuned to be at the same momentum, no HOFA will terminate at that momentum. This is where our classification differs from Ref.~\cite{yang2015topological}: in that classification, two Dirac points can only be inverses if they annihilate each other. In that sense, the classification in Ref.~\cite{yang2015topological} is in momentum space, where a trivial Dirac point implies a band gap, and our classification is in real space, where a trivial Dirac point is one on which a HOFA does not terminate.

We now derive $\Delta \eta$ for each $n$-fold rotation axis.

\begin{table}
\begin{tabular}{|c|c|c|c|c|}
\hline
$n$& Layer group &Symmetry indicator&$\Delta \eta$ &\text{Class.} \\
\hline
$4$&$p4/m'$&$\Delta \eta = \Delta \left( 2[M_\frac12]\right)$& $2 \mod 4$ &${\mathbb Z}_2$\\
$6$&$p6/m'$&$\Delta \eta = \Delta \left( 2[K_\frac12]\right)$& $0,\pm 2 \mod 6$ &${\mathbb Z}_3$\\
$3$&$p\bar{3}$& $\Delta \eta = -\Delta \left( 2[K_\frac12]+2[K'_\frac12]\right)$ &$\pm 2 \mod 6$ &${\mathbb Z}_3$\\
\hline
\end{tabular}
\caption{Classification of Dirac points based on $\Delta \eta$. The first column indicates the $n$-fold rotation that protects the Dirac point (along with $\cal{IT}$ symmetry that squares to $-1$).
The second column indicates the minimal layer group that determines the filling anomaly.
The third column gives the symmetry indicator formula for $\Delta \eta$.
The fourth column gives the possible values of $\Delta \eta$.
When $\Delta \eta \neq 0$ there must be a HOFA terminating  on the Dirac point. When $\Delta \eta =0$, there will not be a HOFA terminating at the Dirac point. This classification is different than Ref.~\cite{yang2015topological} because we adopt a different definition of a trivial phase: in Ref.~\cite{yang2015topological}, the trivial phase is defined as the absence of a Dirac point, while our classification defines a trivial Dirac point as one at which no HOFAs terminate.}
\label{tab:classification}
\end{table}
\par

\subsubsection{$n=4$}
We first list the space groups that have both inversion $\cal I$ and $C_4$ (or four-fold screw) symmetries according to their point groups (with space group number in parenthesis): $4/m$ (no.~83-88), $4/mmm$ (no.~123-142), $m\bar{3}$ (no.~200-206), and $m\bar{3}m$ (no.~221-230). 
If the rod geometry preserves $\cal I$ and $C_4$ symmetries, $p4/m'$ is a subgroup of the layer group of each $k_z$ slice of the rod. 
Therefore, the filling anomaly of each $k_z$ slice is determined by Eq.~(\ref{eqn_eta1}).\par
Dirac points occur when two two-dimensional co-irreps cross, which can happen along the high-symmetry lines $(0,0,k_z)$ or $(\pi,\pi,k_z)$.
These lines are denoted $\Lambda$ or $V$, respectively, in the space group $P4/m$, which we consider as a representative space group without loss of generality. 
The little co-group of both $\Lambda$ and $V$ is $4/m'$, which has only two two-dimensional co-irreps: $E_{\frac12}$ and $E_{\frac32}$. 
These symmetry co-irreps appear in the formula for the filling anomaly (Eq.~(\ref{eqn_eta1})) as $2[M_\frac12]= 2(\# M_\frac12-\# \Gamma_\frac12)$. Since a crossing between the two co-irreps changes $[M_\frac12]$ by $\pm 1$, it results in a change $\Delta\eta^{(4)} = \pm 2$. Since $\eta^{(4)}$ is a mod 4 quantity, $\Delta\eta^{(4)}$ is also defined mod 4. 
Therefore, $\Delta\eta^{(4)} = +2$ is equivalent to $\Delta\eta^{(4)} = -2$. We deduce that there is only one type of Dirac point, which always has a HOFA. This analysis reproduces the result of Ref.~\cite{wieder2020strong}.\par

\subsubsection{$n=6$}
\label{sec:Deltaeta6}
We first list the space groups that have both inversion $\cal I$ and $C_6$ (or six-fold screw) symmetries according to their point groups (with space group number in parenthesis): $6/m$ (no.~175-176), $6/mmm$ (no.~191-194). If the rod geometry preserves $\cal I$ and $C_6$ symmetries, $p6/m'$ is a subgroup of the layer group of each $k_z$ slice of the rod. 
Therefore, the filling anomaly of each $k_z$ slice is determined by Eq.~(\ref{eqn_eta2}).\par

A Dirac point occurs when two two-dimensional co-irreps cross, which can happen along the high-symmetry lines $(0,0,k_z)$ or $(4\pi/3,0,k_z)$, denoted by $\Delta$ or $P$, respectively, in the space group $P6/m$, which we consider as a representative. The little co-group at $\Delta$ is $6/m'$, which has three two-dimensional co-irreps: $E_{\frac12}$, $E_{\frac32}$ and $E_{\frac52}$; the little co-group at $P$ is $\bar{3}'$, which has two two-dimensional co-irreps at $P$: $E_{\frac12}$ and $E_{\frac32}$. We now discuss how crossings between these irreps change the filling anomaly of the 2D $k_z$ slices:
\begin{description}
    \item[Crossings along $\Delta$] According to Eq.~(\ref{eqn_eta2}), the symmetry co-irreps appear in the filling anomaly as $2[K_\frac12]= 2(\# K_\frac12-\# \Gamma_\frac12-\# \Gamma_\frac52)$. Therefore, the crossing between $E_{\frac12}$ and $E_{\frac52}$ does not change the filling anomaly. The crossing between $E_{\frac12}$ and $E_{\frac32}$, and the crossing between $E_{\frac32}$ and $E_{\frac52}$ change the filling anomaly by $\pm 2$.
    \item[Crossings along $P$] There are only crossings between $E_{\frac12}$ and $E_{\frac32}$. The filling anomaly changes by $\pm 2$ according to Eq.~(\ref{eqn_eta2}).
\end{description}
In conclusion, there are three types of Dirac points along $\Delta$ that change the filling anomaly by $\Delta \eta^{(6)}=-2,0$, or $2$ and there are two types of Dirac points along $P$ that change the filling anomaly by $\Delta \eta^{(6)}=\pm 2$.
HOFAs terminate  at Dirac points that have $\Delta\eta \neq 0$.
Thus, not all Dirac points have HOFAs, in contrast to the case of four-fold rotation symmetry.
As we elaborate on in Sec.~\ref{sec:summary}, the presence (absence) of HOFAs corresponds to a linear (quadratic) Dirac point.

\subsubsection{$n=3$}
We first list the space groups that have both inversion $\cal I$ and $C_3$ symmetries according to their point groups (with space group number in parenthesis): $\bar{3}$ (no.~147-148), $\bar{3}m$ (no.~162-167), $m\bar{3}$ (no.~200-206), and $m\bar{3}m$ (no.~221-230). If the rod geometry preserves $\cal I$ and $C_3$ symmetries, $p\bar{3}'$ is a subgroup of the layer group of each $k_z$ slice. 
(We always refer to the $C_3$ axis as the $k_z$ axis.)
The filling anomaly of each $k_z$ slice is defined mod 6 (see Sec.~\ref{sec:Filling_filling}) and determined by Eq.~(\ref{eqn_eta3}).

Dirac points exist when two two-dimensional co-irreps cross along the high-symmetry lines $(0,0,k_z)$ and $(4\pi/3,0,k_z)$, which are denoted by $\Delta$ and $P$ in the representative space group $P\bar{3}$. (Notice this BZ is the same as that of $P6/m$,
but the little co-groups are different; for example, the high symmetry line $(-4\pi/3,0,-k_z)$ is mapped to $(4\pi/3,0,k_z)$ under inversion symmetry.) The little co-group of both $\Delta$ and $P$ is $\bar{3}'$, which has two two-dimensional co-irreps, $E_{\frac12}$ and $E_{\frac32}$.
A crossing between the two co-irreps along either the high symmetry line $\Delta$ and $P$ changes the filling anomaly by $\Delta \eta^{(3)}=\pm 2 \mod 6$.


In conclusion, there are two types of Dirac points along $\Delta$ and $P$ that change the filling anomaly by $\Delta \eta^{(3)}=\pm 2$.
{Since HOFAs terminate  at Dirac points that have $\Delta\eta \neq 0$, all Dirac points have HOFAs.}\par


\subsubsection{Summary of Dirac classification}
\label{sec:summary}

In summary, we have classified Dirac points by how the filling anomaly changes across the Dirac point. 
We have found that for a four-fold rotation axis, there is only one type of Dirac point, corresponding to $\Delta \eta^{(4)}=2\mod 4$; for a six-fold rotation axis, there are three types of Dirac points with $\Delta \eta^{(6)}=0,\pm2\mod 6$; and for a three-fold rotation axis, there are two types of Dirac points with $\Delta \eta^{(3)}=\pm2\mod 6$. These results are summarized in Table~\ref{tab:classification}. 

Since Dirac points have HOFAs terminating on them if and only if $\Delta \eta \neq 0$,
we have now answered the question of when Dirac points have HOFAs: for a Dirac point protected by a three- or four-fold rotation, there is always a HOFA terminating on it, while for a Dirac point protected by a six-fold rotation,
whether or not a HOFA terminates at it can be determined by the symmetry indicator formula in Table~\ref{tab:classification}.

Our analysis is based on symmetry indicators, but we observe that the presence(absence) of a HOFA that terminates at a Dirac point is exactly determined by whether the Dirac point is linear(quadratic).
Specifically, Dirac points protected by three- and four-fold rotations are always linear, while the Dirac point protected by six-fold symmetry and formed by the crossing between the irreps $E_{\frac12}$ and $E_{\frac52}$ is linear in $k_z$ but quadratic in $k_{x,y}$  \cite{yang2014classification}. (The other Dirac points protected by a six-fold rotation are linear \cite{yang2014classification}.)
As we derived in Sec.~\ref{sec:Deltaeta6}, this is the only type of Dirac point that does not have a HOFA terminating on it. 
Thus, we conclude that the linear Dirac points along rotation axes have HOFAs that terminate on them, while the quadratic Dirac points do not.

In a crystal with inversion symmetry,
each Dirac point with change in filling anomaly $\Delta \eta$ has an inversion-partner with $-\Delta\eta$.
The annihilation and creation of Dirac points is compatible with the algebra of $\Delta\eta$. In this sense, the Nielson-Ninomia theorem \cite{nielsen1981absenceI,nielsen1981absenceII} is satisfied and $\Delta\eta$ can be viewed as the topological charge.

For Dirac semimetals with $\cal TI$ symmetry, but not $\cal T$ or $\cal I$ separately, $\Delta\eta$ is still well defined for each Dirac point, and the total topological charge must vanish, but the Dirac points do not have inversion or time-reversed partners. 
When $n=3$ or $n=4$, there are only two irreps that can cross along the high-symmetry lines: thus, due to the periodicity of the BZ, Dirac points must come in pairs (even if there is no symmetry that relates them.)
However, when $n=6$, there are three irreps that can cross and Dirac points need not come in pairs. For example, a band structure of three two-fold degenerate bands with co-irreps $E_\frac12$, $E_\frac32$ and $E_\frac52$ can be arranged so that each pair of co-irreps cross once near the Fermi level, creating three Dirac points near the Fermi level while maintaining the periodicity of the BZ.

\section{\label{sec:TRIM}  TRIM plane topological indices as a classification of Dirac points}

We now compare our classification of Dirac points by the change in the filling anomaly to an earlier classification by Yang and Nagaosa \cite{yang2014classification} where Dirac points are classified by the change in the topological index of TRIM planes (specifically, the mirror Chern number, $C_m$, and time-reversal protected $\mathbb{Z}_2$ invariant, $\nu$).\par

The classification in Ref.~\cite{yang2014classification} was accomplished by constructing $k.p$ models of four band Hamiltonians along the $(0,0,k_z)$ line. 
Here, we use the theory of topological quantum chemistry to systematically reproduce their results, and generalize to situations when Dirac points are along other high symmetry lines.
Our results apply to any number of occupied bands.\par

\begin{table}[]
\begin{tabular}{c|cccc}
$4/m1'$ &$E$&$C_4$ &$C_2$ &$\cal I$\\
\hline
$E_{\frac12 g}$&$2$&$\sqrt{2}$ &$0$ &$2$\\
$E_{\frac12 u}$&$2$&$\sqrt{2}$ &$0$ &$-2$\\
$E_{\frac32 g}$&$2$&$-\sqrt{2}$ &$0$ &$2$\\
$E_{\frac32 u}$&$2$&$-\sqrt{2}$ &$0$ &$-2$
\end{tabular}
\quad
\begin{tabular}{c|ccc}
$2/m1'$ &$E$&$C_2$& $\cal I$\\
\hline
$E_{\frac12 g}$&$2$&$0$ &$2$\\
$E_{\frac12 u}$&$2$&$0$ &$-2$
\end{tabular}
\\
\begin{tabular}{c|cccc}
$6/m1'$ &$E$&$C_6$&$C_3$&$\cal I$\\
\hline
$E_{\frac12 g}$&$2$&$\sqrt{3}$&$1$&$2$\\
$E_{\frac12 u}$&$2$&$\sqrt{3}$&$1$&$-2$\\
$E_{\frac32 g}$&$2$&$0$& $-2$&$2$\\
$E_{\frac32 u}$&$2$&$0$& $-2$&$-2$\\
$E_{\frac52 g}$&$2$&$-\sqrt{3}$&$1$&$2$\\
$E_{\frac52 u}$&$2$&$-\sqrt{3}$&$1$&$-2$
\end{tabular}
\quad
\begin{tabular}{c|ccc}
$\bar{6}1'$ &$E$&$C_3$&$C_6{\cal I}$\\
\hline
$E_{\frac12}$&$2$&$1$ & $\sqrt{3}$\\
$E_{\frac32}$&$2$&$-2$ & $0$\\
$E_{\frac52}$&$2$&$1$ & $-\sqrt{3}$
\end{tabular}
\\
\begin{tabular}{c|ccc}
$\bar{3}1'$ &$E$&$C_3$ &$\cal I$\\
\hline
$E_{\frac12 g}$&$2$&$1$ &$2$ \\
$E_{\frac12 u}$&$2$&$1$ &$-2$ \\
$E_{\frac32 g}$&$2$&$-2$ &$2$\\
$E_{\frac32 u}$&$2$&$-2$ &$-2$
\end{tabular}
\quad 
\begin{tabular}{c|ccc}
$\bar{1}1'$ &$E$ &$\cal I$\\
\hline
$E_{g}$&$2$ &$2$ \\
$E_{u}$&$2$ &$-2$ 
\end{tabular}
\quad
\begin{tabular}{c|ccc}
$31'$ &$E$&$C_3$ \\
\hline
$E_{\frac12 }$&$2$&$1$  \\
$E_{\frac32 }$&$2$&$-2$
\end{tabular}
\caption{Character tables of the spinful irreducible co-representations of the magnetic point groups $4/m1'$, $2/m1'$, $6/m1'$, $\bar{6}1'$, $\bar{3}1'$, $\bar{1}1'$, and $31'$. These point groups are the little co-groups of TRIMs of the magnetic layer groups $p4/m1'$, $p6/m1'$ and $p\bar{3}1'$ as we explain in the text.}
\label{tab_T_irrep}
\end{table}

To this end, we apply the Smith normal form symmetry indicator formula of the stable topological indices \cite{cano2020band,elcoro2020magnetic} to the three minimal magnetic layer groups {at TRIM planes, which are} generated by $\cal T$ (with ${\cal T}^2=-1$), $\cal I$ and $C_n$, where $n=3,4,6$. The details are in Appendix~\ref{app_EBR}. Here we summarize the results:

\begin{description}
\item[$p4/m1'$]
the little co-groups at $\Gamma=(0,0)$, $X=(\pi,0)$, $M=(\pi,\pi)$ are $4/m1'$, $2/m1'$, $4/m1'$ (in the magnetic point group notation). The co-irreps of the little co-groups are listed in Table~\ref{tab_T_irrep}. There is one stable topological index, which is the mirror Chern number:
\begin{equation}
\label{eqn_index1}
    C_m^{(4)} = -[M_{\frac12 g}]+[M_{\frac32 g}]+2[X_{\frac12 u}]-2[M_{\frac12 u}] \mod 4
\end{equation}
where $[X_{\frac12 u}]=\# X_{\frac12 u}-\# \Gamma_{\frac12 u}-\#\Gamma_{\frac32 u}$, $[M_\rho]=\# M_\rho-\#\Gamma_\rho$.
The superscript $4$ in the mirror Chern number $C_m^{(4)}$ indicates the four-fold rotation axis. $C_m^{(4)}$ mod 2 is the 2D strong topological insulator index $\nu$.

\item[$p6/m1'$]
the little co-groups at $\Gamma=(0,0)$, $M=(\pi,-\pi/\sqrt{3})$, $K=(4\pi/3,0)$ are $6/m1'$, $2/m1'$, $\bar{6}1'$. The co-irreps of the little co-groups are listed in Table~\ref{tab_T_irrep}. There is one stable topological index, which is the mirror Chern number:
\begin{equation}
\label{eqn_index2}
    C_m^{(6)} = 2[K_\frac32]+4[K_\frac12]-3[M_{\frac12 g}] \mod 6
\end{equation}
where $[K_\frac12]=\#K_\frac12- \#\Gamma_{\frac12 g}-\#\Gamma_{\frac52 u}$, $[K_\frac32]=\# K_\frac32-\#\Gamma_{\frac32 g}-\#\Gamma_{\frac32 u}$, and $[M_{\frac12g}] = \# M_{\frac12g}-\#\Gamma_{\frac12 g}-\#\Gamma_{\frac32 g}-\#\Gamma_{\frac52 g}$. 
The superscript $6$ in the mirror Chern number $C_m^{(6)}$ indicates the six-fold rotation axis. $C_m^{(6)}$ mod 2 is the 2D strong TI index $\nu$.

\item[$p\bar{3}1'$]
the little co-groups at $\Gamma=(0,0)$, $M=(\pi,-\pi/\sqrt{3})$, $K=(4\pi/3,0)$ are $\bar{3}1'$, $\bar{1}1'$, $31'$. The co-irreps of the little co-groups are listed in Table~\ref{tab_T_irrep}. There is one stable topological index, which is the 2D strong topological insulator index:
\begin{equation}
\label{eqn_index3}
    \nu = [M_u] \mod 2
\end{equation}
where $[M_u]=\# M_u-\# \Gamma_{\frac12 u}-\#\Gamma_{\frac32 u}$. 
\end{description}

\begin{table}[b]
\begin{tabular}{|c|c|c|c|}
\hline
$~n~$  &Classification \cite{yang2014classification}  & Stable TCI index &$~\Delta$index\\
\hline
$4$ & $\mathbb{Z}_4$ &$~\Delta C_m^{(4)}$ (\ref{eqn_index1})& $~\pm 1,~\pm2$  \\
$6$ & $\mathbb{Z}_6$ &$~\Delta C_m^{(6)}$ (\ref{eqn_index2})& $~\pm 1,~\pm 2,~\pm3$  \\
$3$ & $\mathbb{Z}_2$ & $~\Delta \nu$ (\ref{eqn_index3})&$~0,~1\mod 2$ \\
\hline
\end{tabular}
\caption{The classification of Dirac points {with an $n$-fold rotation axis} based on $\Delta C_m$ and $\Delta \nu$. When either is nonzero, there must be gapless surface states on at least one of the TRIM planes. These surface states do not necessitate the existence of Fermi arcs that terminate  at the Dirac points because they can form a closed loop \cite{kargarian2016surface,kargarian2018deformation}.
}
\label{tab:classification_TRIM}
\end{table}
\par

We now can classify the Dirac points by the change in the stable topological index of TRIM planes, i.e. the difference $\Delta C_m$ or $\Delta \nu$ between the $k_z=0$ and $k_z=\pi$ planes. 
Specifically, the high symmetry line where the Dirac point lies contains one TRIM point in each TRIM plane.
The co-representations (co-reps) of the bands that cross at the Dirac point are compatible with specific co-reps at each TRIM point. By plugging the co-rep at each of the two TRIM points into Eq.~(\ref{eqn_index1}), (\ref{eqn_index2}), or (\ref{eqn_index3}), the difference $\Delta C_m$ or $\Delta \nu$ is obtained.
The result is shown in Table~\ref{tab:classification_TRIM}.\par
This classification based on symmetry indicators is in the spirit of Ref.~\cite{yang2014classification}, although goes beyond Ref.~\cite{yang2014classification} by including Dirac points along both high symmetry lines. In particular, the crossings along $P$ in $P6/m$ were not included in the earlier literature. They are distinct from crossings along $\Delta$, because the little co-group at $K$ and $A$ is $\bar{61'}$, while the little co-group at $\Gamma$ and $Z$ is $6/m1'$.
\par

The $k.p$ analysis in Ref.~\cite{yang2014classification} has one advantage {over our classification by topological quantum chemistry}: it distinguishes $\Delta C_m=+ 2$ and $-2$ in the $C_4$ case, and $\Delta C_m=+ 3$ and $-3$ in the $C_6$ case. If one can determine all the parameters of the $k.p$ model and the $k.p$ model is a complete description of the low energy physics, $C_m$ can be determined completely.
In our approach, the $\mathbb{Z}$-valued mirror Chern number can only be partially determined (up to $\mathbb{Z}_n$) by symmetries. 

The changes in topological indices
$\Delta C_m$ and $\Delta \nu$ can also be viewed as topological charges associated with Dirac points (albeit a different charge than the change in filling anomaly). 
Since a Dirac point with $\Delta C_m$ or $\Delta \nu$ must have an inversion-partner with $-\Delta C_m$ or $-\Delta \nu$, the generalized Nielson-Ninomia theorem \cite{nielsen1981absenceI,nielsen1981absenceII} 
is again satisfied.
\par

As a final note on this classification, let us clarify that it does not apply to Dirac cones protected by screw symmetries because the irreps in the $k_z=\pi$ plane are different for a group with a screw symmetry versus a group with a pure rotation symmetry of the same order.
{Since the classification specifically requires symmetry indices in the $k_z = \pi$ plane (unlike our classification based on $\Delta \eta$, which only requires symmetry indices in planes adjacent to the Dirac point), it does not apply.}
 \par

\section{\label{sec:connect}Relation between the filling anomaly and the stable index}
\begin{table*}[]
\begin{tabular}{|c|c|c|c|c|c|c|c|}
\hline \hline
~~n~~ &  k line & co-irreps of Dirac point & dispersion & $\Delta \eta$ & compatible co-irreps at TRIMs  & ~$\Delta C_m$~ &$\Delta \nu \mod 2$  \\ \hline \hline
\multirow{4}{*}{} && \multirow{2}{*}{} & \multirow{2}{*}{} & &$(E_{\frac12 g},~E_{\frac32 g})$~\text{or}~$(E_{\frac12 u},~E_{\frac32 u})$ &  & 0  \\ \cline{6-8} 
                  &$\Delta~(\Gamma A)$&$(E_{\frac12},~E_{\frac32})$ &\text{linear} &~~$\pm 2\mod 6$~~ &~~$(E_{\frac12 g},~E_{\frac32 u})$~\text{or}~$(E_{\frac12 u},~E_{\frac32 g})$~~ &  & 1  \\ \cline{2-8} 
                 3& \multirow{2}{*}{} &&& \multirow{2}{*}{}  &$(E_{\frac12 g},~E_{\frac32 g})$~\text{or}~$(E_{\frac12 u},~E_{\frac32 u})$  &  &0  \\ \cline{6-8} 
                  &$P ~(K H)$ &$(E_\frac12,~E_\frac32)$ &\text{linear} &$\pm 2\mod 6$ &$(E_{\frac12 g},~E_{\frac32 u})$~\text{or}~$(E_{\frac12 u},~E_{\frac32 g})$  &  &1  \\ \hline \hline
\multirow{4}{*}{} & \multirow{2}{*}{} && \multirow{2}{*}{} & &$(E_{\frac12 g},~E_{\frac32 g})$~\text{or}~$(E_{\frac12 u},~E_{\frac32 u})$  & $\pm$ 2  & 0  \\ \cline{6-8} 
                  &$\Lambda~(\Gamma Z)$ &$(E_\frac12,~E_\frac32)$ &\text{linear} & $2\mod 4$& $(E_{\frac12 g},~E_{\frac32 u})$~\text{or}~$(E_{\frac12 u},~E_{\frac32 g})$ & $\pm$ 1 & 1  \\ \cline{2-8} 
                 4& \multirow{2}{*}{} && & \multirow{2}{*}{}  &$(E_{\frac12 g},~E_{\frac32 g})$~\text{or}~$(E_{\frac12 u},~E_{\frac32 u})$  &$\pm 2$ &0 \\ \cline{6-8} 
                  &$V~(M A)$& $(E_\frac12,~E_\frac32)$ &\text{linear}& $2\mod 4$  &$(E_{\frac12 g},~E_{\frac32 u})$~\text{or}~$(E_{\frac12 u},~E_{\frac32 g})$  &$\pm 1$  & 1 \\ \hline \hline
\multirow{8}{*}{} & \multirow{4}{*}{} && & \multirow{2}{*}{} &$(E_{\frac12 g},~E_{\frac32 g})$~\text{or}~$(E_{\frac12 u},~E_{\frac32 u})$  & $\pm$ 2 & 0  \\ \cline{6-8} 
                  & &$(E_\frac12,~E_\frac32)$ &\text{linear} & $\pm 2\mod 6$  &$(E_{\frac12 g},~E_{\frac32 u})$~\text{or}~$(E_{\frac12 u},~E_{\frac32 g})$  & $\pm$ 1 & 1  \\ \cline{3-8} 
                  &  & & \multirow{2}{*}{}  & &$(E_{\frac52 g},~E_{\frac32 g})$~\text{or}~$(E_{\frac52 u},~E_{\frac32 u})$ & $\pm$ 2  & 0  \\ \cline{6-8}
                  &$\Delta ~(\Gamma A)$&$(E_\frac52,~E_\frac32)$ &\text{linear} & $\pm 2\mod 6$ \multirow{2}{*}{} &$(E_{\frac52 g},~E_{\frac32 u})$~\text{or}~$(E_{\frac52 u},~E_{\frac32 g})$ & $\pm$ 1 & 1 \\ \cline{3-8}
                  &  & & \multirow{2}{*}{} & &$(E_{\frac12 g},~E_{\frac52 g})$~\text{or}~$(E_{\frac12 u},~E_{\frac52 u})$ & $\pm$ 2 & 0  \\ \cline{6-8}
                 6&  &$(E_\frac12,~E_\frac52)$ &\text{quadratic} &$0\mod 6$ \multirow{2}{*}{} &$(E_{\frac12 g},~E_{\frac52 u})$~\text{or}~$(E_{\frac12 u},~E_{\frac52 g})$  & $\pm$ 3 & 1  \\ \cline{2-8} 
                  & \multirow{2}{*}{} & & \multirow{2}{*}{} & & $(E_{\frac12 },~E_{\frac32 })$&$\pm 2$ &0  \\ \cline{6-8} 
                  &$P ~(K H)$&$(E_\frac12,~E_\frac32)$ &\text{linear} & $\pm 2\mod 6$ &  $(E_{\frac52 },~E_{\frac32 })$& $\pm 2$ &0 \\ \hline \hline
\end{tabular}
\caption{\label{tab_summary} 
Relations between $\Delta \eta$ and $\Delta C_m$ or $\Delta \nu$.
The first column indicates the $n$-fold rotation symmetry of the crystal.
The second column indicates the high-symmetry line where the Dirac point occurs.
The third column gives possible irreps that can cross along that line to form a Dirac point; the dispersion of that Dirac point is indicated in the fourth column, where linear indicates the Dirac point is linear in all directions and quadratic indicates that it is linear along the axis of rotation and quadratic in other directions.
The fifth column indicates the change in the filling anomaly, $\Delta \eta$, defined in Sec.~\ref{sec:Filling_class}.
Each pair of co-irreps that define a Dirac point is compatible with several different possible co-irreps at TRIMs, as indicated by the sixth column (the irrep notation is explained in Sec.~\ref{sec:connect}). 
The last two columns indicate the possible values of the change in the stable index between the TRIM planes, $\Delta C_m$ and $\Delta \nu$, which are derived in Sec.~\ref{sec:TRIM}.
Tables~\ref{tab_PT_irrep} and~\ref{tab_T_irrep} provide the characters for irreps along high-symmetry lines and at TRIMs, respectively.
}
\end{table*}

A non-zero filling anomaly and a non-zero stable topological index at a TRIM plane give rise to different physical observables:
the former gives rise to HOFAs, while the latter implies gapless surface states at TRIM planes.
In general, it is not possible to determine the filling anomaly from the stable topological index of the TRIM planes and vice versa because the addition of occupied trivial bands with a nontrivial filling anomaly changes the filling anomaly but leaves the stable indices invariant.
\par
Nonetheless, the change in the filling anomaly and the change in the stable index are both constrained by symmetries. The change in the filling anomaly can be determined by the symmetry irreps at the high symmetry lines, while the change in the stable index can be determined by the symmetry irreps at the TRIMs. Compatibility conditions relate these irreps, and thus the change in the filling anomaly and the change in the stable index are related.
\par

We summarize the relations in Table~\ref{tab_summary}. Let us now explain the notation.
Each $C_n$-preserving ($n=3,4,6$) high symmetry line connects two TRIMs.
We denote the occupied co-irreps at the two TRIMs by the pair $(E_{\alpha'},~E_{\beta'})$, where $E_{\alpha'}$ appears at one TRIM point and $E_{\beta'}$ at the other. The indices $\alpha' (\beta')$ each contain two values, $j,\xi$, where $j=\frac12,\frac32,\frac52$ indicates the rotation eigenvalue and $\xi=g,u~(+1,-1)$ indicates the inversion eigenvalue.  For each Dirac point, we need only consider one co-irrep at each TRIM because only one co-irrrep is exchanged at the Dirac point.
The co-irreps at the two TRIM, $(E_{\alpha'}, E_{\beta'})$, uniquely determine the co-irreps $(E_\alpha,~E_\beta)$ at adjacent points on the high symmetry line, where $\alpha, \beta=\frac12,\frac32,\frac52$ indicate the rotation eigenvalues. The definition of these co-irreps are given by the characters in Tables~\ref{tab_PT_irrep} and~\ref{tab_T_irrep}. 
The Dirac point is exactly the crossing between $E_\alpha$ and $E_\beta$ along the high-symmetry line; thus these co-irreps determine the change in the filling anomaly $\Delta\eta$, while the co-irreps $(E_{\alpha'},~E_{\beta'})$ determine the change in the stable index $\Delta C_m$ or $\Delta \nu$.
The mapping $(E_{\alpha'},~E_{\beta'}) \mapsto (E_\alpha,~E_\beta)$ is many to one, which means one value of $\Delta \eta$ is compatible with multiple values of $\Delta C_m$ or $\Delta \nu$, as shown by Table~\ref{tab_summary}. 
Table~\ref{tab_summary} serves to synthesize the results of Sec.~\ref{sec:Filling_class} and Sec.~\ref{sec:TRIM}.

\section{\label{sec:example}Examples}
\begin{figure}
    \centering
    \includegraphics[width=0.8\linewidth]{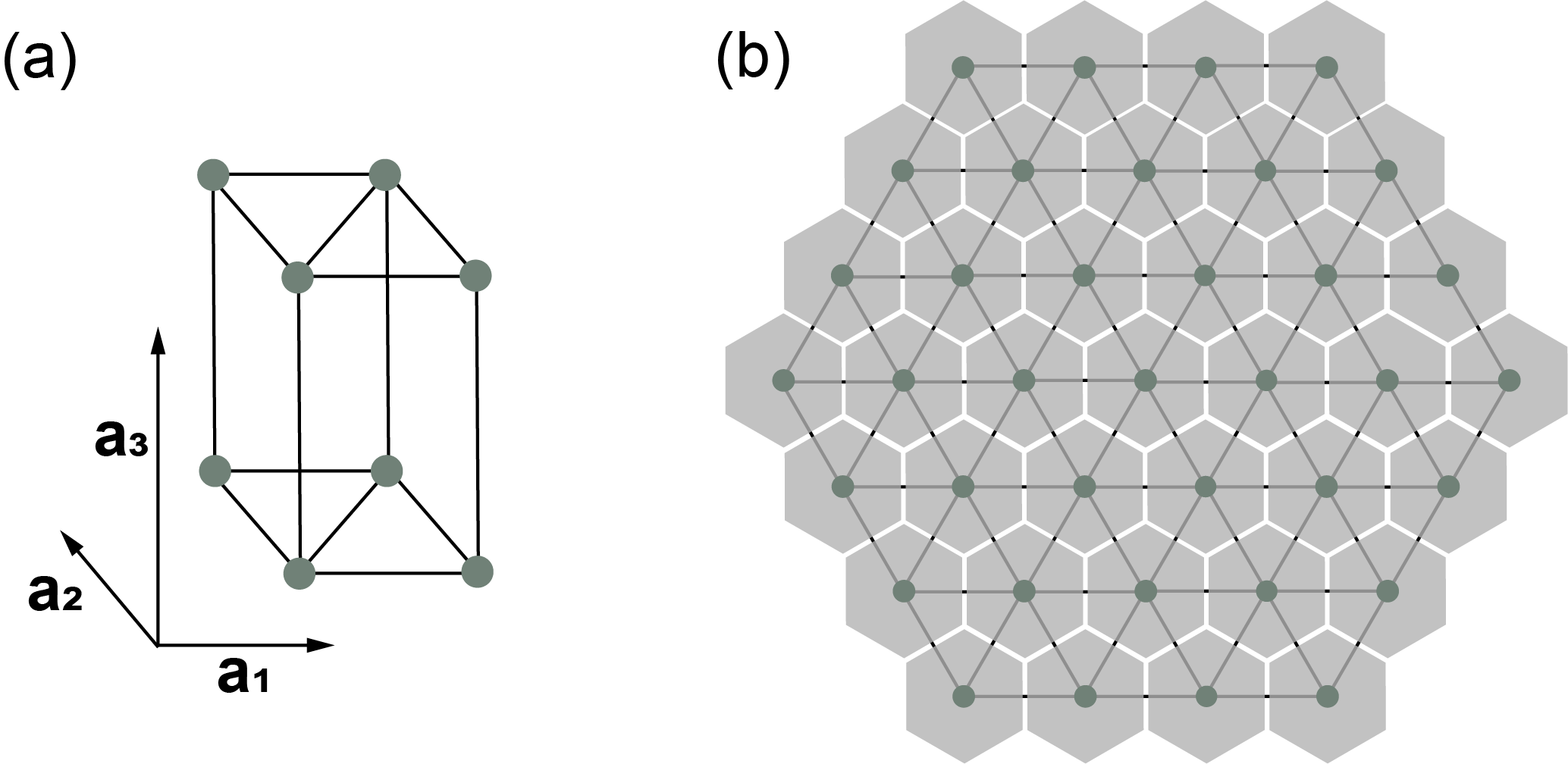}
    \caption{(a) The unit cell of $P6/mmm$. $a_3$ is in the $z$-direction and is perpendicular to $a_1$ and $a_2$. (b) The cross section of a rod which is $C_6$ symmetric, finite in the $a_1$ and $a_2$ directions and infinite in $a_3$ direction.}
    \label{fig_lattice}
\end{figure}

{Ref.~\cite{wieder2020strong} proved that Dirac points in crystals with four-fold rotation symmetry always have HOFAs.}
In this section, we give two examples of Dirac points with six-fold rotation symmetry, with and without HOFAs.
This provides an explicit demonstration that not all Dirac points have HOFAs and verifies the symmetry indicators for $n=6$ in Table~\ref{tab:classification}.

Our models are in space group $P6/mmm$, which has a hexagonal lattice shown in Figure~\ref{fig_lattice}.
In this space group, $\cal TI$ symmetry forces the Hamiltonian to take the form of $H=\epsilon_0(k)+\sum_{j=1}^5 c_j(k)\Gamma_j$, where $k=(k_x,k_y,k_z)$.
We choose a basis where the $\Gamma$ matrices are: 
$$\Gamma_i = \tau_3 \otimes \sigma_i, \quad i=1,2,3$$
$$\Gamma_4 = \tau_2 \otimes \sigma_0, \quad \Gamma_5 = \tau_1 \otimes \sigma_0 $$
where $\sigma_i$ and $\tau_i$, $i=1,2,3$ are Pauli matrices representing the mixed spin and orbital degrees of freedom, $\sigma_0$ is identity matrix and $\otimes$ is the tensor product. 
In this basis, the Hamiltonians for both models take the form \cite{cano2017chiral,wang2012dirac}:
\begin{equation}
\label{eqn_Hamform}
    H=\epsilon_0(k)+\begin{pmatrix}
    {\cal M}(k) & {\cal A}(k) &{\cal B}(k) & 0\\
    {\cal A}^*(k) & -{\cal M}(k) &0 &{\cal B}(k) \\
    {\cal B}^*(k) &0 &-{\cal M}(k) &-{\cal A}(k)\\
    0 & {\cal B}^*(k) &-{\cal A}^*(k) &{\cal M}(k)
     \end{pmatrix}
\end{equation}
The bulk spectrum is $E=\epsilon_0(k)\pm \sqrt{{\cal M}^2+|{\cal A}|^2+|{\cal B}|^2}$. We set $\epsilon_0(k)=0$ for simplicity, which does not influence the topology of the bands. 
The functions ${\cal A}(k)$ and ${\cal B}(k)$ differ between the two models, {but the function ${\cal M}(k)$, which determines the position of Dirac points, is the same.}
We now describe the two models explicitly.\par

\begin{figure}
    \centering
    \includegraphics[width=\linewidth]{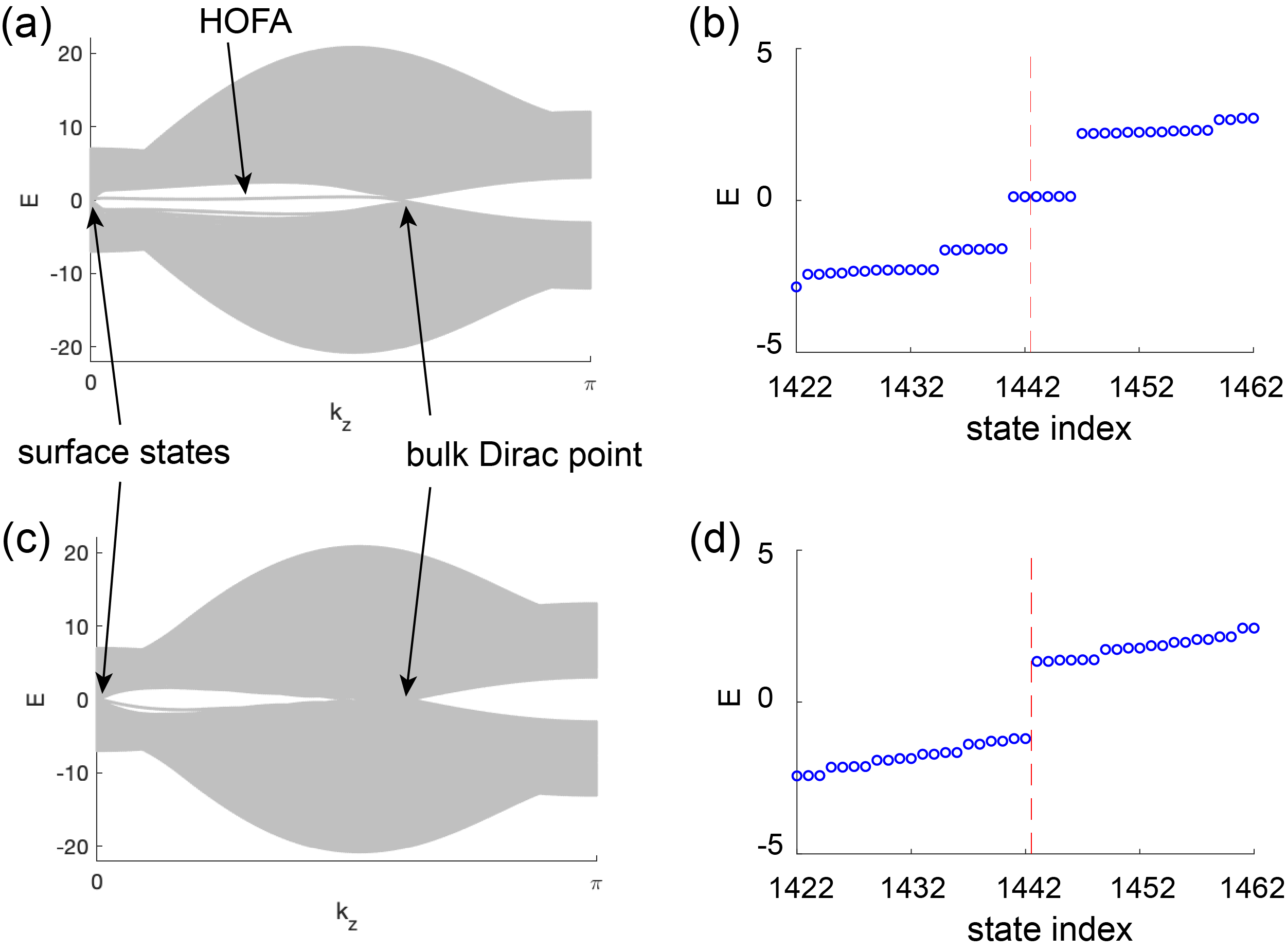}
    \caption{(a) Rod spectrum of the tight-binding model described by Eq.~(\ref{eqn_model1_Ham}) (see Sec.~\ref{sec:example_NaBi} for connection to Na$_3$Bi.) There are HOFAs between $k_z=0$ and $k_z=k_{0}$, the projection of the bulk Dirac point. 
    There are also gapless surface states projected to $k_z=0$. (b) Energy of states at $0<k_z=\pi/4<k_{0}$ for the same model. The dashed red line indicates charge neutrality. 
    {The nontrivial filling anomaly is indicated by the charge neutrality point residing in the middle of six degenerate corner states.}
    (c) Rod spectrum of the tight-binding model described by Eq.~(\ref{eqn_model2_Ham}). There are no HOFA. 
    (d) Energy of states at $0<k_z=\pi/4<k_{0}$ for the second model. The dashed red line indicates charge neutrality. 
    {The lack of filling anomaly is indicated by the charge neutrality point residing in between two groups of degenerate states.}
    There are gapless surface states projecting to $k_z=0$. 
    For both models, the side length of the hexagon cross section (see Figure~\ref{fig_lattice}(b)) is $15$.
    The parameters used to generate the plots are listed in Appendix~\ref{app_models}.}
    \label{fig_spectrum}
\end{figure}

\subsection{\label{sec:example_NaBi} Higher order Fermi arcs in Na$_3$Bi}

Our first example is motivated by the known Dirac semimetal Na$_3$Bi \cite{wang2012dirac,liu2014discovery,xiong2015evidence},
which we predict exhibits higher order Fermi arcs. 
We start in Sec.~\ref{sec:example_NaBi_model} by presenting a simpler toy model on a hexagonal lattice. We use the symmetry indicator formula in Eq.~(\ref{eqn_eta2}) to determine the presence of HOFAs in this model, and verify it with a numerical calculation. Then, in Sec.~\ref{sec:example_NaBi_material}, we discuss the application of our model to the real Na$_3$Bi material.\par

\subsubsection{\label{sec:example_NaBi_model}Model}

In this section, we construct a four-band tight-binding model on a hexagonal lattice in space group $P6/mmm$. 
We first introduce the basis and symmetry of the orbitals and then derive a $k\cdot p$ Hamiltonian to third order.
We construct a tight-binding model by extending the $k\cdot p$ model to the whole Brillouin zone. Finally, we use the symmetry indicator formula in Eq.~(\ref{eqn_eta1}) to study the HOFAs and verify the formulas by plotting the tight-binding spectrum on a rod geometry that is finite in two dimensions.\par

The orbital notation follows Ref.~\cite{wang2012dirac}: $|S^+_{\frac12}, j_z\rangle$ denotes $s$-orbitals with angular momentum quantum numbers $(J,j_z)=(\frac12,\pm \frac12)$ and inversion eigenvalue $+1$, while $|P^-_{\frac32}, j_z\rangle$ denotes $p$-orbitals with $(J,j_z)=(\frac32,j_z)$, $j_z=\pm \frac12,\pm \frac32$, 
and inversion eigenvalue $-1$.\par
We work in the four-band basis: $|P^-_{\frac32}, +\frac32\rangle$, $|S^+_{\frac12}, +\frac12\rangle$, $|S^+_{\frac12}, -\frac12\rangle$, $|P^-_{\frac32}, -\frac32\rangle$; these are the relevant orbitals in Na$_3$Bi~\cite{wang2012dirac}.
In this basis, the symmetry generators of $6/mmm$ ($D_{6h}$), which is the little group at $\Gamma$, are 
\begin{equation}
    C_{6z}=\begin{pmatrix}
    -i &0 &0 &0\\
    0& e^{-i\pi/6} &0 &0\\
    0& 0& e^{i\pi/6} & 0\\
    0&0 &0 & i
    \end{pmatrix}
\end{equation}
\begin{equation}
    C_{2x}=\begin{pmatrix}
    0&0 &0 &i\\
    0&0 & -i&0\\
    0&-i &0  &0 \\
    i&0 &0 &0 
    \end{pmatrix}
\end{equation}
\begin{equation}
    {\cal I}=\begin{pmatrix}
    -1 &0 &0 &0\\
    0& 1 &0 &0\\
    0&0 & 1 &0 \\
    0&0 & 0& -1
    \end{pmatrix}
\end{equation}
and time-reversal symmetry is implemented by
\begin{equation}
    {\cal T} = \begin{pmatrix}
    0 &0 &0 &-1\\
    0&  0&-1 &0\\
    0&1 &0  &0 \\
    1&0 &0 &0 
    \end{pmatrix}K
\end{equation}
where $K$ represents the complex conjugation operator. In the third order $k.p$ approximation, the coefficients in Eq.~(\ref{eqn_Hamform}) are:
\begin{align}
    {\cal M}(k) &= m+M_1k_z^2+M_2k_+k_- \nonumber \\
    {\cal A}(k) &= Ak_-(1+A_1k_z^2+A_2k_+k_-) \nonumber\\
    {\cal B}(k) &= Bk_-^2k_z
    \label{eqn_MAB1}
\end{align}
where $k_\pm=k_x\pm ik_y$ and all the parameters are real. \par

The Dirac point appears along the high symmetry line $\Delta=(0,0,k_z)$, along which ${\cal A, B}=0$, and $\cal M$ changes sign across the Dirac point. The co-irreps in the valence bands that swap across the Dirac point are $E_\frac12$ and $E_\frac32$.
According to Eq.~(\ref{eqn_eta2}), this crossing changes the filling anomaly by $\Delta \eta^{(6)}=\pm 2\mod 6$.
The non-zero change $\Delta \eta^{(6)}$ implies that HOFAs terminate  at the Dirac points.

We would like to extend this $k.p$ Hamiltonian to a tight-binding Hamiltonian to explicitly verify the presence of hinge arcs. 
We now describe the tight-binding model.
The hexagonal lattice vectors are $a_1=(1,0,0)$, $a_2=(-\frac12,\frac{\sqrt{3}}{2},0)$,  $a_3=(0,0,1)$. The tight-binding model is defined by the {real space} hopping terms:
\begin{equation}
    V_{a_1} = \tilde{M_2}\Gamma_3+i\tilde{A}\Gamma_1, \,\,
    V_{a_1+a_3} = \tilde{B}\Gamma_5, \,\,
    V_{a_3} = \tilde{M_1}\Gamma_3
\end{equation}
and an onsite term $m\Gamma_3$, where $V_r$ indicates the hopping matrix from an atom at $r_0$ to an atom at $r_0+r$. 
Other hopping terms are related to these terms by $C_{6z}$ and $C_{2x}$ symmetries. Parameters with a tilde are real and are proportional to the corresponding $k.p$ parameters in Eq.~(\ref{eqn_MAB1}).

The tight-binding Hamiltonian in momentum space takes the form of Eq.~(\ref{eqn_Hamform}), with coefficients:
\begin{align}
    {\cal M}(k) &= m+2 \tilde{M_1} \cos k_z +4 \tilde{M_2} \cos \frac{k_x}{2} \cos \frac{\sqrt{3} k_y}{2} \nonumber\\
    & +2 \tilde{M_2} \cos k_x \nonumber\\
    {\cal A}(k) &=-2 \tilde{A} {\Big( }  \sin \frac{k_x}{2} \cos \frac{\sqrt{3} k_y}{2}+i\sqrt{3} \cos \frac{k_x}{2} \sin \frac{\sqrt{3} k_y}{2} \nonumber\\
    &+\sin k_x {\Big)}\nonumber\\
    {\cal B}(k) &= 2 \tilde{B} \sin k_z {\Big( }\cos k_x-\cos \frac{k_x}{2} \cos \frac{\sqrt{3} k_y}{2} \nonumber\\
    &- i \sqrt{3} \sin \frac{k_x}{2} \sin \frac{\sqrt{3} k_y}{2}{\Big) }
\label{eqn_model1_Ham}
\end{align}
Dirac points can exist along the two high symmetry lines $\Delta$ and $P$, where ${\cal}A(k)={\cal B}(k)=0$. The term ${\cal M}(k)$ determines whether and where there is a Dirac point: if $|m+6\tilde{M}_2|<|2\tilde{M}_1|$ there is a pair of Dirac points along $\Delta$, while if $|m-3\tilde{M}_2|<|2\tilde{M}_1|$, there is a pair of Dirac points along $P$.
{In the interest of connecting to Na$_3$Bi, we choose $\tilde{M_2}=-1$, $m=4$, $\tilde{M_1}=-2.5$, so that there is exactly a pair of Dirac points along $\Delta$, related by inversion symmetry. (Although in Na$_3$Bi, the Dirac points are much closer to $\Gamma$; here, we choose parameters where the Dirac points are further from $\Gamma$ so that the HOFAs are clearly visible.)}
In Figure~\ref{fig_spectrum}(a), we plot the spectrum of this model in a rod geometry (the rod cross-section is shown in Figure~\ref{fig_lattice}(b)), {which shows the existence of mid-gap HOFAs terminating  on the Dirac point.}
Figure~\ref{fig_spectrum}(b) shows that the filling anomaly at $k_z=\pi/4$ (a representative $k_z$-slice with corner states) is $2\mod 6$, in agreement with Eq.~(\ref{eqn_eta2}).
The bulk and surface band structures are plotted in Appendix~\ref{app_models}. \par

There are also surface cones projected onto $k_z=0$ in the rod spectrum in Figure~\ref{fig_spectrum}(a). Such states are often seen in Dirac semimetal systems with HOFAs in the presence of both $\cal T$ and $\cal I$ symmetries because the degeneracy of mid-gap states is $6$ at non-TRIM planes, but $12$ at TRIMs. 
The gapless surface states ensure the rod band structure is continuous where the HOFAs pass through a TRIM plane.
\par

\subsubsection{\label{sec:example_NaBi_material}Application to Na$_3$Bi}

In this section, we discuss the connection between the model described in the previous section and the Dirac material Na$_3$Bi.
Na$_3$Bi is in the non-symmorphic space group $P6_3/mmc$, while our tight-binding model described by Eq.~(\ref{eqn_model1_Ham}) is in the symmorphic space group $P6/mmm$;
the six-fold rotation symmetry in the latter is replaced by a six-fold screw symmetry in the former.
Although the little co-groups of the high-symmetry points are different for the two space groups, the co-irreps of the two groups are the same, up to an overall phase factor, for momenta not in the $k_z = \pi$ plane \cite{elcoro2020magnetic}. {(In the $k_z = \pi$ plane, inversion does not commute with the screw symmetry; this is not an issue for the other planes because when $k_z = 0$, inversion commutes with the screw symmetry, and when $0 < k_z < \pi$, inversion does not leave the plane invariant. For a general discussion of Dirac points protected by screw symmetries, see Appendix~\ref{app:DP_snp}.)}
Therefore, when $k_z\neq \pi$, our model in Eq.~(\ref{eqn_model1_Ham}) in space group $P6/mmm$ applies also to space group $P6_3/mmc$.
Since the low-energy physics in Na$_3$Bi occurs near $\Gamma$, this model also describes the low-energy physics in Na$_3$Bi and in fact the $k.p$ model in Eq.~(\ref{eqn_MAB1}) is identical to that in Ref.~\cite{wang2012dirac}. 

From the perspective of topology, the HOFAs exist in $k_z$ planes where the filling anomaly $\eta$ is non-zero. Eq.~(\ref{eqn_eta2}) implies that $\eta$ depends on the charge of atoms. 
Our four-band Hamiltonians in Eq.~(\ref{eqn_MAB1}) and Eq.~(\ref{eqn_model1_Ham}) have four orbitals coming from a single atom in the unit cell, while, because Na$_3$Bi is in a non-symmorphic space group, it necessarily has multiple symmetry-related Na atoms in the unit cell \cite{wang2012dirac}. Thus, our model does not map in real space to a Hamiltonian in the non-symmorphic space group. 
However, whether HOFAs terminate  at the Dirac points does not depend on $\eta$, but only $\Delta \eta$, i.e., the difference in $\eta$ across the Dirac point, and $\Delta \eta$ does not depend on the atomic positions, only on the co-reps of the bands that cross to form the Dirac point.
Thus, from our model, we predict that Na$_3$Bi has HOFAs that terminate  at the Dirac point, {but we cannot predict whether they cross the $k_z = 0$ or $k_z = \pi$ plane.}
\par

The possible 2D stable topological indices at $k_z=0$ are the same for both space groups (they both have a 2D strong topological insulator index and a mirror Chern number). 
In both cases, a band inversion at $\Gamma$ that creates a pair of Dirac points along the $k_z$ axis will change the 2D stable topological index of the $k_z=0$ plane \cite{yang2014classification}, which is captured by our $C_m$ (mirror Chern) index in Eq.~(\ref{eqn_index2}). 
Since the low-energy physics in Na$_3$Bi is exactly due to such a band inversion, we expect the 2D topological indices in the $k_z = 0$ plane to agree between the two models.

We note that a non-trivial bulk invariant \cite{tyner2020topology} has also been computed for Na$_3$Bi in Ref.~\cite{tyner2021quantized}, but the corner states were not computed.

\subsection{\label{sec:example_noFA} Absence of higher order Fermi arc}
{We now describe a model of a Dirac semimetal with $C_6$ symmetry which does not exhibit HOFAs.}
We work in the four-band basis $|D^+_{\frac52}, +\frac52\rangle$, $|P^-_{\frac32}, +\frac12\rangle$, $|P^-_{\frac32}, -\frac12\rangle$, $|D^+_{\frac52}, -\frac52\rangle$, where $|D^+_{\frac52}, j_z\rangle$ denotes $d$-orbitals with $(J,j_z)=(\frac52,j_z)$, $j_z=\pm \frac12,\pm \frac32,\pm\frac52$. In this basis the symmetry generators of space group $P6/mmm$ are 
\begin{equation}
    C_{6z}=\begin{pmatrix}
    e^{-i5\pi/6} &0 &0 &0\\
    0& e^{-i\pi/6} &0 &0\\
    0&0 & e^{i\pi/6} &0 \\
    0& 0&0 & e^{i5\pi/6}
    \end{pmatrix}
\end{equation}
\begin{equation}
    C_{2x}=\begin{pmatrix}
    0& 0& 0&-i\\
    0& 0& i&0\\
    0&i &0  &0 \\
    -i&0 &0 &0 
    \end{pmatrix}
\end{equation}
\begin{equation}
    {\cal I}=\begin{pmatrix}
    1 &0 &0 &0\\
    0& -1 &0 &0\\
    0&0 & -1 &0 \\
    0&0 & 0& 1
    \end{pmatrix}
\end{equation}
and time-reversal symmetry is implemented by
\begin{equation}
    {\cal T} = \begin{pmatrix}
    0 &0 &0 &-1\\
    0& 0 &1 &0\\
    0&-1 &0  &0 \\
    1&0 &0 &0 
    \end{pmatrix}K
\end{equation}
where $K$ represents the complex conjugation operator. To third order, the coefficients of the $k.p$ Hamiltonian in Eq.~(\ref{eqn_Hamform}) are 
\begin{align}
    {\cal M}(k) &= m+M_1k_z^2+M_2k_+k_- \nonumber \\
    {\cal A}(k) &= Ak_-^2k_z \nonumber\\
    {\cal B}(k) &= B_1k_x(k_x^2-3k_y^2)+iB_2k_y(k_y^2-3k_x^2)
    \label{eqn_MAB2}
\end{align}
where $k_\pm=k_x\pm ik_y$. 

The Dirac point along $\Delta$ results from a crossing between the co-irreps $E_{\frac12}$ and $E_{\frac52}$. According to Eq.~(\ref{eqn_eta2}), this crossing changes the filling anomaly by $\Delta \eta^{(6)}=0 \mod 6$, which implies there is no HOFA that terminates at the Dirac point. Thus, there are two possibilities: (i) no HOFA for any $k_z$; (ii) HOFAs for every $k_z$.  
To determine which possibility occurs, we need a real space embedding.
To this end, we construct a tight binding model to explicitly present an example of (i), i.e., a Dirac point without HOFAs.\par

This $k.p$ Hamiltonian can be extended (non-uniquely) to the whole BZ by a tight-binding model with real space hopping terms:
\begin{align}
    V_{a_1} = \tilde{M_2}\Gamma_3+i\tilde{B_1}\Gamma_5, \quad
    V_{a_1+2a_2} = i\tilde{B_2}\Gamma_4, \nonumber \\
    V_{a_1+a_3} = \tilde{A}\Gamma_1, ~\qquad
    V_{a_3} = \tilde{M_1}\Gamma_3
\end{align}
and the onsite term $m\Gamma_3$. Other hopping terms are related to these terms by $C_{6z}$ and $C_{2x}$ symmetries. Parameters with a tilde are real and are proportional to the corresponding $k.p$ parameters {in Eq.~(\ref{eqn_MAB2})}.
\par
The tight-binding Hamiltonian in momentum space is of the form of Eq.~(\ref{eqn_Hamform}), with coefficients:
\begin{align}
    {\cal M}(k) & = m+2 \tilde{M_1} \cos k_z+4 \tilde{M_2} \cos \frac{k_x}{2} \cos \frac{\sqrt{3} k_y}{2} \nonumber\\
    &+2 \tilde{M_2} \cos k_x \nonumber\\
    {\cal A}(k) &= 2 \tilde{A} \sin k_z {\Big( }\cos k_x-\cos \frac{k_x}{2} \cos \frac{\sqrt{3} k_y}{2} \nonumber\\
    &- i \sqrt{3} \sin \frac{k_x}{2} \sin \frac{\sqrt{3} k_y}{2}{\Big) } \nonumber\\
    {\cal B}(k) &= 4 \tilde{B_1} \sin \frac{k_x}{2} \cos \frac{\sqrt{3} k_y}{2}-2 \tilde{B_1}\sin k_x \nonumber\\
    &-4 i \tilde{B_2} \cos \frac{3 x}{2} \sin \frac{\sqrt{3} k_y}{2}+2 i \tilde{B_2} \sin \sqrt{3} k_y
    \label{eqn_model2_Ham}
\end{align}

Similar to the previous model, the term ${\cal M}(k)$ determines whether and where there is a Dirac point: if $|m+6\tilde{M}_2|<|2\tilde{M}_1|$ there is a pair of Dirac points along $\Delta$, while if $|m-3\tilde{M}_2|<|2\tilde{M}_1|$, there is a pair of Dirac points along $P$. 
{We consider $\tilde{M_2}=-1$, $m=4$, $\tilde{M_1}=-2.5$, so that again there is exactly a pair of Dirac points along $\Delta$, related by inversion symmetry.}
In Figure~\ref{fig_spectrum}(c), we plot the rod spectrum, {which shows the absence of HOFAs}. Figure~\ref{fig_spectrum}(d) shows that the filling anomaly at $k_z=\pi/4$ (a representative $k_z$-slice between the Dirac point and $k_z=0$) is $0\mod 6$, which is consistent with the absence of HOFAs in Figure~\ref{fig_spectrum}(c).
\par

In Appendix~\ref{app_models}, we describe more details about both models, including the parameters used to plot the band structures, the bulk and surface BZs, the bulk and surface spectra, and the band co-representations, which verify the presence/absence of HOFAs with symmetry indicators.\par

\section{Discussion}

{Dirac fermions protected by inversion, time-reversal, and four-fold rotation symmetry have been shown to exhibit HOFAs \cite{wieder2020strong,lin2018topological}, which are mid-gap states localized at the corners where two surfaces meet that connect the projection of the bulk Dirac points. The HOFAs can be viewed as a topological bulk-hinge correspondence for Dirac fermions.
In this manuscript, we derived conditions under which HOFAs exist for any Dirac fermion protected by inversion, time-reversal, and an $n$-fold rotation symmetry, where $n=3,4,6$. }
We further computed symmetry indicators to determine the presence/absence of HOFAs, using an algorithm we introduced in Ref.~\cite{fang2021filling}.

We found that all linear Dirac points exhibit HOFAs that terminate on them. 
However, a six-fold rotation symmetry can also protect a Dirac point that has a linear dispersion along the rotation axis and a quadratic dispersion in the other directions; this Dirac point does not exhibit HOFAs.


Our results define a topological charge for each Dirac point based on the change in the filling anomaly, $\Delta \eta$, of 2D planes in momentum space on either side of the Dirac point.
We found that in the case of a four-fold rotation, $\Delta \eta^{(4)}=2\mod 4$, corresponding to a single type of Dirac point;
in the case of a six-fold rotation, $\Delta \eta^{(6)}=0,\pm2\mod 6$, corresponding to three types of Dirac points;
and finally in the case of a three-fold rotation, $\Delta \eta^{(3)}=\pm2\mod 6$, corresponding to two types of Dirac points.

We built explicit tight-binding models in the case of a six-fold rotation symmetry to exemplify Dirac semimetals with and without HOFAs. \par

Our formulas also apply to Dirac points protected by screw symmetries if they are not on the boundary of the BZ.
The bulk-hinge correspondence of Dirac points at the BZ boundary is a subject for future work.


\begin{acknowledgements}
J.C. and Y.F. acknowledge a useful conversation with Barry Bradlyn and acknowledge Simin Nie for pointing out a typo in the manuscript.
J.C. acknowledges the support of the Flatiron Institute, a division of Simons Foundation.
This material is based upon work supported by the National Science Foundation under Grant No. DMR-1942447.

\end{acknowledgements}

\bibliography{reference}

\begin{appendix}
\section{\label{app:DP} {Symmetry of} Dirac points}

In this Appendix, we discuss the symmetries that can protect Dirac points.
In \ref{app:DP_Cn}, we derive that only rotation symmetries of order 3, 4, or 6 can protect Dirac points.
In \ref{app:DP_snp}, we derive that screw symmetries of the same order can also protect Dirac points; only two of these are consistent with inversion symmetry.
In \ref{sec_DP_summary}, we derive the centrosymmetric space groups that contain the rotation and screw symmetries that can protect Dirac points.
We briefly discuss other symmetries that can protect Dirac points on the boundary of the BZ in Appendix~\ref{sec_DP_discuss}.\par

\subsection{\label{app:DP_Cn}Dirac points stabilized by rotation symmetry}
Dirac points not at a TRIM are protected by the combination of $\cal TI$ symmetry and an $n$-fold rotation.
These symmetries generate the groups $2/m'$, $\bar{3}'$, $4/m'$, and $6/m'$ for $n=2,3,4,6$, which are the only rotations that occur in crystals.
We only consider the case with spin-orbit coupling, where ${\cal TI}$ is an antiunitary symmetry that squares to $-1$ and commutes with all rotations;
thus, it requires all {bands} to be doubly degenerate.
Since for $n=2$, there is only one two-dimensional co-irrep, a two-fold rotation symmetry cannot protect a Dirac point (because the Dirac point requires two different co-irreps to cross).
For $n=3$ and $n=4$, there are two two-dimensional co-irreps and hence there are two choices of crossings.
For $n=6$, there are three two-dimensional co-irreps and, consequently, there are six choices of crossings.
We enumerate the co-irreps of $\bar{3}'$, $4/m'$, and $6/m'$ in Table~\ref{tab_PT_irrep}, where the notation follows Ref.~\cite{altmann1994point}.\par

A Dirac material may also have additional mirror symmetries.
Mirror symmetries that leave the rotation axis invariant enlarge its little co-group to one of the following magnetic point groups: $\bar{3}'m$, $4/m'mm$, or $6/m'mm$.
In each of these groups, the number of two-dimensional co-irreps with SOC is the same as the number without the additional mirror symmetries \cite{elcoro2020magnetic}.
Thus, adding the mirror symmetry does not change the types of Dirac crossings.
We do not consider mirror symmetries that invert the high-symmetry line because they will not leave generic points along the high-symmetry line invariant.

\subsection{\label{app:DP_snp}Screw symmetries}

In this subsection, we show that screw symmetries can also protect Dirac points.

\begin{figure*}
    \centering
    \includegraphics[width=0.9\linewidth]{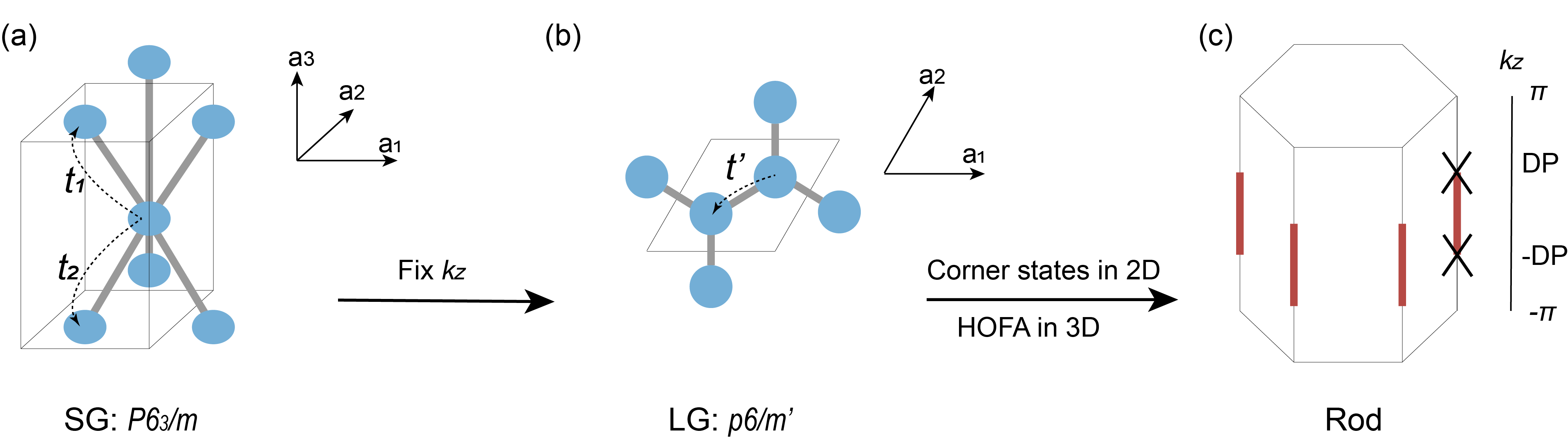}
    \caption{In this schematic diagram, we take space group $P6_3/m$ (with ${\cal T}^2=-1$) as an example of a case with screw symmetry. (a) Hexagonal unit cell of space group $P6_3/m$. (b) A fixed $k_z$ slice with $k_z\neq 0,\pi$ forms an effective 2D system described by the magnetic layer group $p6/m'$. The atoms project to $z=0$ in the unit cell of this 2D system. The interlayer hopping term $t'=t_1e^{ik_z/2}+t_2 e^{-ik_z/2}$ {becomes an in-plane hopping term}. (c) Terminating the 2D models with boundaries that preserve $C_6$ symmetry, yields the spectrum of the rod states for the 3D model. Red lines indicate possible HOFAs that could appear between two Dirac points (whose projection onto the hinges is labelled by crosses).}
    \label{fig_scheme2}
\end{figure*}

{In momentum space, a screw symmetry can be represented as the product of a $k$-independent unitary matrix and a $k$-dependent phase \cite{alexandradinata2016topological}. 
Specifically, a screw symmetry that rotates by an angle $2\pi/n$ and translates by a fractionn $p/n$ of a lattice vector can be represented by the matrix} $\tilde{C}_{n,p}=e^{-ik_z\frac pn}U_n$ in momentum space, where $n=2,3,4,6$; $p$ is an integer satisfying $1\leq p\leq n-1$; and $U_n$ is a unitary matrix where
$U_n^n = -1$ is equivalent to a $2\pi$ rotation.
The screw symmetry constrains the Hamiltonian by \cite{tsirkin2017composite} 
\begin{equation}
    \tilde{C}_{n,p} H(\mathbf{k})\tilde{C}_{n,p}^{-1} = U_n H(\mathbf{k}) U_n^{-1} =H(R_n\mathbf{k}),
\end{equation}
where $R_n$ is the vector representation of the $n$-fold rotation symmetry. 
Notice the translation by $p/n$ has dropped out of this constraint.
Thus, in momentum space, a screw symmetry acts effectively as a rotation symmetry. 
{It follows that screw symmetries can protect a Dirac crossing if $n=3,4,6$.}

If $\cal I$ symmetry is present, then the translation $p$ must satisfy $2p=n$ because the combined operation $\tilde{C}_{n,p}^{-1}{\cal I}\tilde{C}_{n,p}{\cal I}:(x,y,z)\rightarrow (x,y,z-2p/n)$ must give a lattice translation, which implies $-2p/n \in \mathbb{Z}$. Since $1\leq p\leq n-1$, the only solution is $2p=n$.
Therefore, with the combination of time-reversal and inversion symmetry, only $\tilde{C}_{4,2}$ and $\tilde{C}_{6,3}$ can protect Dirac points that are not at the TRIM points.

For these reasons, in 3D crystal systems with a screw symmetry $\tilde{C}_{n,p}=e^{-ik_z\frac pn}U_n$, the effective 2D system with fixed $k_z\neq \pi$, is described by the unitary part $U_n$. 
Therefore, when $k_z\neq \pi$ the filling anomaly formulas Eq.~(\ref{eqn_eta1}) and Eq.~(\ref{eqn_eta2}) also apply to $\tilde{C}_{4,2}$ and $\tilde{C}_{6,3}$.
An example is shown in Figure~\ref{fig_scheme2}. Here the dimensional reduction of a 3D model with the space group $P6_3/m$ is an effective 2D model, described by the magnetic layer group $p6/m'$. Then the rod geometry should preserve six-fold rotation symmetry. 

Such dimensional reduction can break down if there are other symmetries whose commutation relation with the screw symmetry $\tilde{C}_{n,p}$ is different from the commutation relation with its rotational part $U_n$. In the simplest case where the only symmetries are the screw rotation, inversion, and time-reversal symmetry, such break down appears at high symmetry points with $k_z=\pi$ because, as Refs.~\cite{elcoro2017double,yang2015topological} show, the screw symmetry anti-commutes with inversion symmetry at $k_z=\pi$, while rotation symmetry commutes with inversion symmetry at $k_z=\pi$. This anti-commutation relation gives rise to the four- and higher-dimensional co-irreps at the boundary of BZ. Those Dirac points are beyond the scope of this manuscript. We provide some known examples in Appendix~\ref{sec_DP_discuss} for completeness. 



\subsection{{Space groups with Dirac points}\label{sec_DP_summary}}
We have shown that in the presence of $\cal T$ and $\cal I$, where $(\mathcal{TI})^2 = -1$, Dirac points not at TRIM points are stabilized by $C_n$, $n=3,4,6$, or $\tilde{C}_{4,2}$ or $\tilde{C}_{6,3}$.
The space groups containing these symmetries must have a point group containing $\cal I$ and $C_n$, $n=3,4,6$.
There are in total 8 point groups (58 space groups) that have the required symmetries.  We list the point groups here (with corresponding space group numbers in parenthesis): $4/m$ (no.~83-88), $4/mmm$ (no.~123-142), $\bar{3}$ (no.~147, no.~148), $\bar{3}m$ (no.~162-167), $6/m$ (no.~175, no.~176), $6/mmm$ (no.~191-194), $m\bar{3}$ (no.~200-206), $m\bar{3}m$ (no.~221-230). \par
Notice that {since a Dirac point requires only the combination $\cal TI$, rather than $\cal T$ and $\cal I$ separately,} there will also be Dirac points in magnetic space groups that have $\cal TI$ symmetry and one of the necessary rotation/screw symmetries. The minimal magnetic space groups are subgroups of the space groups we consider in the main text. Thus, our classification of Dirac points by the change in filling anomaly also applies to those magnetic groups.\par

\subsection{Dirac points on the BZ boundary \label{sec_DP_discuss}}

There are other Dirac points in non-symmorphic space groups that fall outside of our paradigm. 
These Dirac points are all at the boundary of the BZ. The reason why the boundary of the BZ is special is because in the interior of the BZ, the representation of the little group is always the same as the representation of a point group, which is a subgroup of the factor group, but on the boundaries of the BZ, the representations of the non-symmorphic symmetry operations will be the projective representations of the point group, which differ from the ordinary representation at momenta that are in the interior of the BZ \cite{bradley2009mathematical}. We now briefly describe some examples.

Screw symmetries (combined with $\cal TI$ symmetry) can protect a single Dirac point at a boundary TRIM, as we discussed in Appendix~\ref{app:DP_snp}. These Dirac points are linear in $k_z$, since they are created by band folding \cite{young2012dirac}. 
The bands that cross to form the Dirac point have different eigenvalues of the screw symmetry.
\par

Two-fold screw symmetries can also protect Dirac points that are not at TRIMs but are located at the boundary of the BZ. One example is shown in Ref.~\cite{wieder2016double}. Although Ref.~\cite{wieder2016double} is focusing on the double Dirac point at a boundary TRIM, there are also Dirac points on a boundary high symmetry line protected by two non-symmorphic symmetries in space group $P4/ncc$ (No. 130).

Glide symmetries can also protect Dirac points that are not at TRIMs but are located at the boundary of the BZ \cite{yang2015topological}. 
\par

There can also exist quadratic Dirac points at TRIMs, such as in $Fm\bar{3}m$. These also fall outside our paradigm.
In space group $Fm\bar{3}m$, there can exist a single Dirac point at $\Gamma$ in the BZ. This Dirac point is quadratic in all directions. The four bands forming the degeneracy comprise a single irrep of the point group.

\section{\label{app:wyck} Unit cells and Wyckoff positions}

\begin{figure}
    \centering
    \includegraphics[width = 0.8\linewidth]{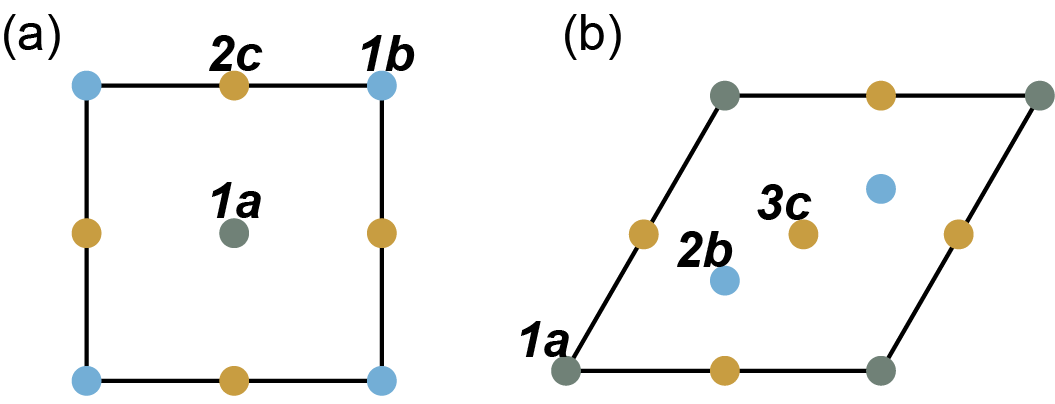}
    \caption{The projected 2D unit cell and Wyckoff positions of (a) $p4/m'$ and $p4/m1'$ and (b) $p\bar{3}$, $p6/m'$, $p\bar{3}1'$, and $p6/m1'$.}
    \label{fig_wyckoff}
\end{figure}

The projected 2D unit cells and the maximal Wyckoff positions of the magnetic layer groups $p4/m'$, $p6/m'$, $p\bar{3}'$ and $p4/m1'$, $p6/m1'$, $p\bar{3}1'$ are shown in Figure~\ref{fig_wyckoff}. The general Wyckoff positions are not shown because they do not contribute to the filling anomaly or the stable topological index \cite{bradlyn2017topological,cano2020band,fang2021filling}. 
We now summarize the maximal Wyckoff positions and corresponding site symmetry groups {when one of these 2D groups describes a momentum-space slice of a 3D rod, as depicted in Figure~\ref{fig_scheme}}. 

For $p4/m'$ (Figure~\ref{fig_wyckoff}(a)), there are three maximal 
Wyckoff positions: $1a$, $1b$ and $2c$. The magnetic site symmetry group of $1a$ and $1b$ is $4/m'$; the magnetic site symmetry group of $2c$ is $2/m'$ (generated by $C_2$ and $\cal TI$).\par
For $p6/m'$ (Figure~\ref{fig_wyckoff}(b)), there are three maximal 
Wyckoff positions: $1a$, $2b$ and $3c$. The magnetic site symmetry group of $1a$ is $6/m'$; The magnetic site symmetry group of $2b$ is $\bar{6}'$ (generated by $C_6{\cal TI}$); The magnetic site symmetry group of $3c$ is $2/m'$.\par
For $p\bar{3}'$ (Figure~\ref{fig_wyckoff}(b)), there are three maximal 
Wyckoff positions: $1a$, $2b$ and $3c$. The magnetic site symmetry group of $1a$ is $\bar{3}'$; The magnetic site symmetry group of $2b$ is $3$ (generated by $C_3$); The magnetic site symmetry group of $3c$ is $\bar{1}'$ (generated by $\cal TI$).\par

For $p4/m1'$ (Figure~\ref{fig_wyckoff}(a)), there are three maximal 
Wyckoff positions: $1a$, $1b$ and $2c$. The magnetic site symmetry group of $1a$ and $1b$ is $4/m1'$; the magnetic site symmetry group of $2c$ is $2/m1'$ (generated by $C_2$, $\cal I$ and $\cal T$).\par
For $p6/m1'$ (Figure~\ref{fig_wyckoff}(b)), there are three maximal 
Wyckoff positions: $1a$, $2b$ and $3c$. The magnetic site symmetry group of $1a$ is $6/m1'$; The magnetic site symmetry group of $2b$ is $\bar{6}1'$ (generated by $C_6{\cal I}$ and ${\cal T}$); The magnetic site symmetry group of $3c$ is $2/m1'$.\par
For $p\bar{3}1'$ (Figure~\ref{fig_wyckoff}(b)), there are three maximal 
Wyckoff positions: $1a$, $2b$ and $3c$. The magnetic site symmetry group of $1a$ is $\bar{3}1'$; The magnetic site symmetry group of $2b$ is $31'$ (generated by $C_3$ and $\cal T$); The magnetic site symmetry group of $3c$ is $\bar{1}1'$ (generated by $\cal I$ and $\cal T$).\par

We have listed the minimal magnetic layer groups that we consider in the main text. 
The character tables for site-symmetry groups utilized are in Tables~\ref{tab_PT_irrep} and \ref{tab_T_irrep}. \par
Adding mirror symmetries may change the Wyckoff positions,
but the symmetry indicators for the filling anomaly (Eqs.~(\ref{eqn_eta1}), (\ref{eqn_eta2}) and~(\ref{eqn_eta3}) 
and the symmetry indices for the stable topological phases (Eqs.~(\ref{eqn_index1}),  (\ref{eqn_index2}) and (\ref{eqn_index3})) that we derived from the groups without mirror symmetries still apply. 
This is because the minimal magnetic layer groups are subgroups of the groups with mirror symmetries and, importantly, the irreps of the minimal groups are the same as the irreps of the groups with mirror symmetry \cite{elcoro2017double}. Further, we can always project the Wyckoff positions invariant under mirror symmetries to Wyckoff positions in the groups without mirror symmetries. After this projection, the symmetry indicator {formulas} are valid.

\section{\label{app:r} {Derivation of modulus of $\eta$}}
In this section, we use the representation theory of magnetic groups to derive the modulus, $r$, of the filling anomaly in Eq.~(\ref{eqn_eta_nw}).
Specifically, we prove that when the symmetry group contains $\mathcal{TI}$ but not $\cal T$ or $\cal I$ separately,
the modulus of {$\eta$ in} Eq.~(\ref{eqn_eta_nw}) is $r$, while when $\cal T$ and $\cal I$ symmetries are both present, the modulus of $\eta$ is $2r$. 
If the crystal has four-fold rotation symmetry, $r=4$, while if it has three- or six-fold rotation symmetry, $r=6$.\par

We take the two dimensional finite lattice to be a regular polygon that preserves the rotation and inversion symmetries, i.e., a square when $n=4$ and a hexagon when $n=3$ or $6$.
The argument can be generalized to other symmetric lattices.
We denote the (magnetic) point group of the finite lattice by $G$ and then explain how $G$ completely determines the modulus of the filling anomaly.

The modulus of filling anomaly is the least number of electrons that can be added to the boundary of the finite crystal without breaking symmetry. To preserve symmetry, the electrons must be added to a Wyckoff position of the (magnetic) point group $G$. 
{We define a Wyckoff position of a point group to be a set of points whose site-symmetry groups are conjugate; here the site-symmetry group of a particular site is defined to be the subgroup of the point group that leaves that site invariant.
For the group $C_n$, the Wyckoff positions of the point group include the $1a$ position at the rotation center, whose site-symmetry group is $G$, and one non-maximal general position, whose site-symmetry group is the identity.
For the group $C_{nv}$, which has additional mirror planes, there is an additional Wyckoff position in each mirror plane, whose site-symmetry group is generated by the mirror symmetry (for example, in $C_{2v}$, there is a $2b$ position containing $(x,0)$ and $(-x,0)$, invariant under $m_y$ and a $2c$ position containing $(0,y)$ and $(0,-y)$, invariant under $m_x$.)}

Since the modulus of the filling anomaly is the least number of electrons that can be added to the boundary of the finite crystal while preserving symmetry, those electrons can be added to any Wyckoff position except the $1a$ position (because the $1a$ position is not at the boundary.) Let $\mathbf{q}$ be the Wyckoff position of smallest multiplicity besides the $1a$ position.
Then the modulus of $\eta$ is determined by the product of the multiplicity of $\mathbf{q}$ and the dimension of the smallest (co-)irrep of the (magnetic) site-symmetry group of $\mathbf{q}$.
We now describe how to find the dimension of this (co-)irrep.

When $\cal T$ symmetry is present, the smallest co-irrep is always {two-dimensional} as required by the Kramers degeneracy. The Wyckoff positions and representative coordinates for $p\bar{3}'$, $p4/m'$, and $p6/m'$ are $6b~(x,y,z)$, $4b~(x,y,0)$, and $6b~(x,y,0)$, respectively. 
Therefore, the modulus of filling anomaly is $12$, $8$, {and} $12$ respectively.

When $\cal TI$ symmetry is present, {but not $\mathcal{T}$,} the Wyckoff positions are the same as the $\cal T$ symmetric cases. Therefore, the choice of $\mathbf q$ is the same. But the smallest co-irrep is always one-dimensional because there is not an anti-unitary symmetry that squares to $-1$ in the site-symmetry group. Specifically, for $p\bar{3}'$ there is no anti-unitary symmetry that leaves $\mathbf q$ invariant; for $p4/m'$ and $p6/m'$, $C_2{\cal TI}$ is an anti-unitary symmetry that leaves $\mathbf q$ invariant, but it squares to $+1$. 
Thus, in this case, for $n=3,4,6$ the modulus of filling anomaly is $6$, $4$, $6$ respectively. 

\section{\label{app_EBR}EBR analysis}
The analysis of elementary band representations in this work is based on the (magnetic) topological quantum chemistry theory established in Refs.~\cite{bradlyn2017topological,cano2018building,elcoro2020magnetic}. For a pedagogical review, we refer the readers to Ref.~\cite{cano2020band}. We now summarize the essential points.\par

Let $A$ be the integer ``EBR matrix'' of the symmetry group under consideration: each column of $A$ is labelled by an EBR and each row a particular irrep of the little group of a particular high-symmetry point.
The entries in the matrix indicate the number of times each momentum space irrep appears in the EBR \cite{cano2020band,song2020fragile,song2020twisted}.
A group of topologically trivial bands can be expressed as a linear combination of EBRs \cite{bradlyn2017topological} with integer coefficients ${n}_i$.
The irreps that appear at high-symmetry points in the band structure satisfy
\begin{equation}
    v=A{n},
    \label{eqn_vAn}
\end{equation}
where $v_j$ is the number of times the $j^\text{th}$ irrep appears in the band structure. \par
Let the Smith normal form of $A$ be given by
\begin{equation}
    A = U^{-1}DV^{-1},
    \label{eqn_smithA}
\end{equation}
where $D$ is a diagonal positive integer matrix with diagonal entries $(d_1, \dots, d_M, 0, \dots 0)$, i.e., the first $M$ entries are positive and the remaining entries are zero, and $U,V$ are integer matrices invertible over the integers.
The stable topological classification is given by
\begin{equation}
\label{eqn_Zdm}
    \mathbb{Z}_{d_1} \times \cdots \times \mathbb{Z}_{d_M}.
\end{equation} 
{Notice that if $d_m=1$ for some $1\leq m\leq M$, then the $m^{\rm th}$ component of the classification is trivial (because $\mathbb{Z}_1$ has only one element.)}
Therefore, the classification is determined by those $d_m>1$.

{We now describe how to determine the topological index of a group of bands whose symmetry co-reps at high symmetry points are given by $v$; 
we say that $v$ has a nontrivial stable index if no integer solution to Eq.~(\ref{eqn_vAn}) exists and has a trivial stable index otherwise.}
The $m^{\text{th}}$ component of the stable topological index (Eq.~(\ref{eqn_Zdm})) of the group of bands labelled by $v$ is given by \cite{po2017symmetry,song2018quantitative,song2020fragile,song2020twisted,cano2020band}:
\begin{equation}
    \text{index} = \left( Uv \right)_m \mod d_m
    \label{eqn_index}
\end{equation}
where $1\leq m\leq M$, and $d_m>1$. 
If the index of a given band representation vanishes and the system is in the trivial phase (i.e. in the atomic limit), then the valence bands admit a Wannier representation. The number of Wannier functions that have Wannier centers at the maximal Wyckoff position $w$ can be determined by the following formula as discussed in Ref.~\cite{fang2021filling}:
\begin{multline}
    e_w =  \sum_{i\in w} \text{dim}(\rho_i) \left[ VD^pUv\right]_i  \\
    \mod  \text{gcd}\lbrace \left( \sum_{i\in w} \text{dim}(\rho_i) V_{ij} \right) | _{j > M} \rbrace.
    \label{eqn_smithew}
\end{multline}
The sum over ${i\in w}$ means the sum over EBRs that are induced from a representation $\rho_i$ of the site symmetry group of the Wyckoff position $w$. $D^p$ is the pseudo-inverse of $D$, which is a diagonal matrix with diagonal entries $(d_1^{-1},...,d_M^{-1},0,...,0)$. gcd means the greatest common divisor.
This formula also applies to the fragile phase where at least one component of $n$ is a negative integer \cite{fang2021filling,benalcazar2019quantization}.
\par
In the following subsections, we present the EBR matrix $A$ and Smith decomposition of the magnetic groups that we use in the main text. The EBR matrices are derived from the Bilbao Crystallographic Server~\cite{elcoro2020magnetic}.
Plugging the matrices $D$, $U$ and $V$ into Eq.~(\ref{eqn_index}) and Eq.~(\ref{eqn_smithew}), we determine the symmetry indicator formulas for the stable topological indices in Eqs.~(\ref{eqn_index1}), (\ref{eqn_index2}), (\ref{eqn_index3}), and the filling anomalies in Eqs.~(\ref{eqn_eta1}), (\ref{eqn_eta2}) and (\ref{eqn_eta3}).

\subsection{\label{app_EBR_4} $p4/m'$}

The basis for band co-representations (columns) and the basis for coefficients of EBRs (rows) are
\begin{equation}
    \left( E^{\Gamma}_{\frac12},E^{\Gamma}_{\frac32},E^{X}_{\frac12},E^{M}_{\frac12},E^{M}_{\frac32} \right),
\end{equation}
{and}
\begin{equation}
    \left( E^{1a}_{\frac12},E^{1a}_{\frac32},E^{1b}_{\frac12},E^{1b}_{\frac32},E^{2c}_{\frac12} \right).
\end{equation}
In these bases, the EBR matrix is
\begin{equation}
    \label{eqn_4_A}
    A = \begin{pmatrix}
    1 & 0 & 1 & 0 & 1 \\
 0 & 1 & 0 & 1 & 1 \\
 1 & 1 & 1 & 1 & 2 \\
 1 & 0 & 0 & 1 & 1 \\
 0 & 1 & 1 & 0 & 1 \\
    \end{pmatrix}
\end{equation}
The Smith normal form matrices are
\begin{equation}
    \label{eqn_4_D}
    D = \left(
\begin{array}{ccccc}
 1 & 0 & 0 & 0 & 0 \\
 0 & 1 & 0 & 0 & 0 \\
 0 & 0 & 1 & 0 & 0 \\
 0 & 0 & 0 & 0 & 0 \\
 0 & 0 & 0 & 0 & 0 \\
\end{array}
\right)
\end{equation}

\begin{equation}
    \label{eqn_4_U}
    U = \begin{pmatrix}
    0 & 0 & 0 & 1 & 0 \\
 0 & 1 & 0 & 0 & 0 \\
 1 & 0 & 0 & -1 & 0 \\
 -1 & -1 & 1 & 0 & 0 \\
 -1 & -1 & 0 & 1 & 1 \\
    \end{pmatrix}
\end{equation}

\begin{equation}
    \label{eqn_4_V}
    V = \begin{pmatrix}
    1 & 0 & 0 & -1 & -1 \\
 0 & 1 & 0 & -1 & -1 \\
 0 & 0 & 1 & 1 & 0 \\
 0 & 0 & 0 & 1 & 0 \\
 0 & 0 & 0 & 0 & 1 \\
    \end{pmatrix}
\end{equation}
\par
Since all the non-zero elements in the diagonal matrix $D$ (Eq.~(\ref{eqn_4_D})) are $1$, there is no symmetry indicated stable topological index {per Eq.~(\ref{eqn_Zdm})}.
Plugging these matrices into Eq.~(\ref{eqn_smithew}), we get Eq.~(\ref{eqn_na1}) and Eq.~(\ref{eqn_nb1}) in Sec.~\ref{sec:Filling_indicator}.

\subsection{\label{app_EBR_6} $p6/m'$}
The basis for band co-representations (columns) and the basis for coefficients of EBRs (rows) are
\begin{equation}
    \left( E^{\Gamma}_{\frac32},E^{\Gamma}_{\frac12},E^{X}_{\frac52},E^{K}_{\frac32},E^{K}_{\frac12} \right),
\end{equation}
{and}
\begin{equation}
    \left( E^{1a}_{\frac32},E^{1a}_{\frac12},E^{1a}_{\frac52},E^{2b}_{\frac32},E^{2b}_{\frac12} \oplus E^{2b}_{\frac52} \right).
\end{equation}
In these bases, the EBR matrix is
\begin{equation}
    \label{eqn6_A}
    A = \left(
\begin{array}{ccccc}
 1 & 0 & 0 & 1 & 0 \\
 0 & 1 & 0 & 0 & 1 \\
 0 & 0 & 1 & 0 & 1 \\
 1 & 0 & 0 & 0 & 1 \\
 0 & 1 & 1 & 1 & 1 \\
\end{array}
\right)
\end{equation}
The Smith normal form matrices are 
\begin{equation}
    \label{eqn_6_D}
    D = \left(
\begin{array}{ccccc}
 1 & 0 & 0 & 0 & 0 \\
 0 & 1 & 0 & 0 & 0 \\
 0 & 0 & 1 & 0 & 0 \\
 0 & 0 & 0 & 1 & 0 \\
 0 & 0 & 0 & 0 & 0 \\
\end{array}
\right)
\end{equation}

\begin{equation}
    \label{eqn_6_U}
    U = \left(
\begin{array}{ccccc}
 0 & 0 & 0 & 1 & 0 \\
 0 & 1 & 0 & 0 & 0 \\
 0 & 0 & 1 & 0 & 0 \\
 1 & 0 & 0 & -1 & 0 \\
 -1 & -1 & -1 & 1 & 1 \\
\end{array}
\right)
\end{equation}

\begin{equation}
    \label{eqn_6_V}
    V = \left(
\begin{array}{ccccc}
 1 & 0 & 0 & 0 & -1 \\
 0 & 1 & 0 & 0 & -1 \\
 0 & 0 & 1 & 0 & -1 \\
 0 & 0 & 0 & 1 & 1 \\
 0 & 0 & 0 & 0 & 1 \\
\end{array}
\right)
\end{equation}
\par
Since all the non-zero elements in the diagonal matrix $D$ (Eq.~(\ref{eqn_6_D})) are $1$, there is no symmetry indicated stable topological index {per Eq.~(\ref{eqn_Zdm})}.
Plugging these matrices into Eq.~(\ref{eqn_smithew}), we get Eq.~(\ref{eqn_na2}) and Eq.~(\ref{eqn_nb2}) in Sec.~\ref{sec:Filling_indicator}.

\subsection{\label{app_EBR_3} $p\bar{3}'$}
The basis for band co-representations (columns) and the basis for coefficients of EBRs (rows) are
\begin{equation}
    \left( E^{\Gamma}_{\frac12}, E^{\Gamma}_{\frac32}, E^{K}_{\frac12}, E^{K}_{\frac32}, E^{K'}_{\frac12}, E^{K'}_{\frac32} \right),
\end{equation}
{and}
\begin{equation}
    \left( E^{1a}_{\frac12},E^{1a}_{\frac32}, {}^1E^{2b}_{\frac12}, {}^2E^{2b}_{\frac12}, E^{2b}_{\frac32} \right).
\end{equation}
In these bases, the EBR matrix is
\begin{equation}
    \label{eqn_3_A}
    A = \left(
\begin{array}{ccccc}
1 & 0 & 1 & 1 & 0 \\
0 & 1 & 0 & 0 & 1 \\
1 & 0 & 1 & 0 & 1 \\
0 & 1 & 0 & 1 & 0 \\
1 & 0 & 0 & 1 & 1 \\
0 & 1 & 1 & 0 & 0 \\
\end{array}
\right)
\end{equation}
The Smith normal form matrices are
\begin{equation}
    \label{eqn_3_D}
    D = \left(
\begin{array}{ccccc}
 1 & 0 & 0 & 0 & 0 \\
 0 & 1 & 0 & 0 & 0 \\
 0 & 0 & 1 & 0 & 0 \\
 0 & 0 & 0 & 1 & 0 \\
 0 & 0 & 0 & 0 & 0 \\
 0 & 0 & 0 & 0 & 0 \\
\end{array}
\right)
\end{equation}

\begin{equation}
    \label{eqn_3_U}
    U = \left(
\begin{array}{cccccc}
 0 & 1 & 0 & -1 & 1 & 0 \\
 0 & 1 & 0 & 0 & 0 & 0 \\
 1 & 0 & 0 & 0 & -1 & 0 \\
 0 & -1 & 0 & 1 & 0 & 0 \\
 -1 & -1 & 1 & 1 & 0 & 0 \\
 -1 & -1 & 0 & 0 & 1 & 1 \\
\end{array}
\right)
\end{equation}

\begin{equation}
    \label{eqn_3_V}
    V =\left(
\begin{array}{ccccc}
 1 & 0 & 0 & 0 & -2 \\
 0 & 1 & 0 & 0 & -1 \\
 0 & 0 & 1 & 0 & 1 \\
 0 & 0 & 0 & 1 & 1 \\
 0 & 0 & 0 & 0 & 1 \\
\end{array}
\right)
\end{equation}
\par
Since all the non-zero elements in the diagonal matrix $D$ (Eq.~(\ref{eqn_3_D})) are $1$, there is no symmetry indicated stable topological index {per Eq.~(\ref{eqn_Zdm})}. Plugging these matrices into Eq.~(\ref{eqn_smithew}), we get Eq.~(\ref{eqn_eta3}) and Eq.~(\ref{eqn_nb3}) in Sec.~\ref{sec:Filling_indicator}.

\subsection{\label{app_EBR_41} $p4/m1'$}
The basis for band co-representations (columns) and the basis for coefficients of EBRs (rows) are
\begin{equation}
\label{eqn_basis_br_4}
    \left( E^{\Gamma}_{\frac12g}, E^{\Gamma}_{\frac12u}, E^{\Gamma}_{\frac32g}, E^{\Gamma}_{\frac32u}, E^{X}_{g}, E^{X}_{u}, E^{M}_{\frac12g}, E^{M}_{\frac12u}, E^{M}_{\frac32g}, E^{M}_{\frac32u} \right) ,
\end{equation}
{and}
\begin{equation}
\label{eqn_basis_ebr_4}
    \left( E^{1a}_{\frac12g}, E^{1a}_{\frac12u}, E^{1a}_{\frac32g}, E^{1a}_{\frac32u}, E^{1b}_{\frac12g}, E^{1b}_{\frac12u}, E^{1b}_{\frac32g}, E^{1b}_{\frac32u}, E^{2c}_{g}, E^{2c}_{u} \right) .
\end{equation}
In these bases, the EBR matrix is
\begin{equation}
    \label{eqn_41_A}
   A= \begin{pmatrix}
    1 & 0 & 0 & 0 & 1 & 0 & 0 & 0 & 1 & 0 \\
 0 & 1 & 0 & 0 & 0 & 1 & 0 & 0 & 0 & 1 \\
 0 & 0 & 1 & 0 & 0 & 0 & 1 & 0 & 1 & 0 \\
 0 & 0 & 0 & 1 & 0 & 0 & 0 & 1 & 0 & 1 \\
 1 & 0 & 1 & 0 & 0 & 1 & 0 & 1 & 1 & 1 \\
 0 & 1 & 0 & 1 & 1 & 0 & 1 & 0 & 1 & 1 \\
 1 & 0 & 0 & 0 & 0 & 0 & 1 & 0 & 0 & 1 \\
 0 & 1 & 0 & 0 & 0 & 0 & 0 & 1 & 1 & 0 \\
 0 & 0 & 1 & 0 & 1 & 0 & 0 & 0 & 0 & 1 \\
 0 & 0 & 0 & 1 & 0 & 1 & 0 & 0 & 1 & 0 \\
    \end{pmatrix}
\end{equation}
The Smith normal form matrices are
\begin{equation}
    \label{eqn_41_D}
    D = \begin{pmatrix}
    1 & 0 & 0 & 0 & 0 & 0 & 0 & 0 & 0 & 0 \\
 0 & 1 & 0 & 0 & 0 & 0 & 0 & 0 & 0 & 0 \\
 0 & 0 & 1 & 0 & 0 & 0 & 0 & 0 & 0 & 0 \\
 0 & 0 & 0 & 1 & 0 & 0 & 0 & 0 & 0 & 0 \\
 0 & 0 & 0 & 0 & 1 & 0 & 0 & 0 & 0 & 0 \\
 0 & 0 & 0 & 0 & 0 & 1 & 0 & 0 & 0 & 0 \\
 0 & 0 & 0 & 0 & 0 & 0 & 1 & 0 & 0 & 0 \\
 0 & 0 & 0 & 0 & 0 & 0 & 0 & 4 & 0 & 0 \\
 0 & 0 & 0 & 0 & 0 & 0 & 0 & 0 & 0 & 0 \\
 0 & 0 & 0 & 0 & 0 & 0 & 0 & 0 & 0 & 0 \\
    \end{pmatrix}
\end{equation}

\begin{equation}
    \label{eqn_41_U}
    U = \begin{pmatrix}
    0 & 0 & 0 & 0 & 0 & 0 & 1 & 0 & 0 & 0 \\
 0 & 0 & 0 & 0 & 0 & 0 & 0 & 1 & 0 & 0 \\
 0 & 0 & 1 & 0 & 0 & 0 & 0 & 0 & 0 & 0 \\
 0 & 0 & 0 & 1 & 0 & 0 & 0 & 0 & 0 & 0 \\
 0 & 0 & 0 & -1 & 0 & 1 & 0 & -1 & 0 & 0 \\
 1 & 1 & 1 & 0 & 0 & 0 & -1 & -1 & -1 & 0 \\
 0 & 0 & 1 & -1 & 0 & 1 & 0 & -1 & -1 & 0 \\
 1 & 0 & -1 & 2 & 0 & -2 & -1 & 2 & 1 & 0 \\
 -1 & -1 & -1 & -1 & 1 & 1 & 0 & 0 & 0 & 0 \\
 -1 & -1 & -1 & -1 & 0 & 0 & 1 & 1 & 1 & 1 \\
    \end{pmatrix}
\end{equation}

\begin{equation}
    \label{eqn_41_V}
    V = \begin{pmatrix}
    1 & 0 & 0 & 0 & 0 & 0 & 0 & 1 & -1 & -1 \\
 0 & 1 & 0 & 0 & 0 & 0 & -1 & -2 & -1 & -1 \\
 0 & 0 & 1 & 0 & 0 & 0 & -1 & -1 & -1 & -1 \\
 0 & 0 & 0 & 1 & 0 & 0 & 0 & 0 & -1 & -1 \\
 0 & 0 & 0 & 0 & 1 & 0 & 0 & 1 & 1 & 0 \\
 0 & 0 & 0 & 0 & 0 & 1 & -1 & -2 & 1 & 0 \\
 0 & 0 & 0 & 0 & 0 & 0 & 0 & -1 & 1 & 0 \\
 0 & 0 & 0 & 0 & 0 & 0 & 0 & 0 & 1 & 0 \\
 0 & 0 & 0 & 0 & 0 & 0 & 1 & 2 & 0 & 1 \\
 0 & 0 & 0 & 0 & 0 & 0 & 0 & 0 & 0 & 1 \\
    \end{pmatrix}
\end{equation}

The {diagonal entry of 4 in the} matrix $D$ (Eq.~(\ref{eqn_41_D})) shows there is one symmetry indicated stable topological index {per Eq.~(\ref{eqn_index})}.
Plugging these matrices into Eq.~(\ref{eqn_index}), we get Eq.~(\ref{eqn_index1}) in Sec.~\ref{sec:TRIM}. It is a mirror Chern number, determined by symmetries mod 4.

\subsection{\label{app_EBR_61} $p6/m1'$}
The basis for band co-representations (columns) and the basis for coefficients of EBRs (rows) are
\begin{equation}
\label{eqn_basis_br_6}
    \left( E^{\Gamma}_{\frac32g}, E^{\Gamma}_{\frac32u}, E^{\Gamma}_{\frac12g}, E^{\Gamma}_{\frac12u}, E^{\Gamma}_{\frac52g}, E^{\Gamma}_{\frac52u}, E^{X}_{g}, E^{X}_{u}, E^{M}_{\frac32}, E^{M}_{\frac12}, E^{M}_{\frac52} \right) ,
\end{equation}
{and}
\begin{equation}
\label{eqn_basis_ebr_6}
    \left( E^{1a}_{\frac32g}, E^{1a}_{\frac32u}, E^{1a}_{\frac12g}, E^{1a}_{\frac12u}, E^{1a}_{\frac52g}, E^{1a}_{\frac52u}, E^{2b}_{\frac32}, E^{2b}_{\frac12}, E^{2b}_{\frac52}, E^{3c}_{g}, E^{3c}_{u} \right) .
\end{equation}
In these bases, the EBR matrix is
\begin{equation}
    \label{eqn_61_A}
A =\left( \begin{array}{ccccccccccc}
 1 & 0 & 0 & 0 & 0 & 0 & 1 & 0 & 0 & 1 & 0 \\
 0 & 1 & 0 & 0 & 0 & 0 & 1 & 0 & 0 & 0 & 1 \\
 0 & 0 & 1 & 0 & 0 & 0 & 0 & 1 & 0 & 1 & 0 \\
 0 & 0 & 0 & 1 & 0 & 0 & 0 & 0 & 1 & 0 & 1 \\
 0 & 0 & 0 & 0 & 1 & 0 & 0 & 0 & 1 & 1 & 0 \\
 0 & 0 & 0 & 0 & 0 & 1 & 0 & 1 & 0 & 0 & 1 \\
 1 & 0 & 1 & 0 & 1 & 0 & 1 & 1 & 1 & 1 & 2 \\
 0 & 1 & 0 & 1 & 0 & 1 & 1 & 1 & 1 & 2 & 1 \\
 1 & 1 & 0 & 0 & 0 & 0 & 0 & 1 & 1 & 1 & 1 \\
 0 & 0 & 1 & 0 & 0 & 1 & 1 & 0 & 1 & 1 & 1 \\
 0 & 0 & 0 & 1 & 1 & 0 & 1 & 1 & 0 & 1 & 1 
\end{array}
\right)
\end{equation}
The Smith normal form matrices are
\begin{equation}
    \label{eqn_61_D}
    D = \left(\begin{array}{ccccccccccc}
    1 & 0 & 0 & 0 & 0 & 0 & 0 & 0 & 0 & 0 & 0 \\
 0 & 1 & 0 & 0 & 0 & 0 & 0 & 0 & 0 & 0 & 0 \\
 0 & 0 & 1 & 0 & 0 & 0 & 0 & 0 & 0 & 0 & 0 \\
 0 & 0 & 0 & 1 & 0 & 0 & 0 & 0 & 0 & 0 & 0 \\
 0 & 0 & 0 & 0 & 1 & 0 & 0 & 0 & 0 & 0 & 0 \\
 0 & 0 & 0 & 0 & 0 & 1 & 0 & 0 & 0 & 0 & 0 \\
 0 & 0 & 0 & 0 & 0 & 0 & 1 & 0 & 0 & 0 & 0 \\
 0 & 0 & 0 & 0 & 0 & 0 & 0 & 1 & 0 & 0 & 0 \\
 0 & 0 & 0 & 0 & 0 & 0 & 0 & 0 & 6 & 0 & 0 \\
 0 & 0 & 0 & 0 & 0 & 0 & 0 & 0 & 0 & 0 & 0 \\
 0 & 0 & 0 & 0 & 0 & 0 & 0 & 0 & 0 & 0 & 0 \\
    \end{array}
    \right)
\end{equation}

\begin{equation}
    \label{eqn_61_U}
    U =\left( \begin{array}{ccccccccccc}
    1 & 0 & 1 & 0 & 0 & 1 & 0 & 0 & 0 & -1 & 0 \\
 0 & 1 & 1 & 0 & 0 & 1 & 0 & 0 & 0 & -1 & 0 \\
 0 & 0 & 1 & 0 & 0 & 0 & 0 & 0 & 0 & 0 & 0 \\
 0 & 0 & 0 & 1 & 0 & 0 & 0 & 0 & 0 & 0 & 0 \\
 0 & 0 & 0 & 0 & 1 & 0 & 0 & 0 & 0 & 0 & 0 \\
 0 & 0 & 0 & 0 & 0 & 1 & 0 & 0 & 0 & 0 & 0 \\
 1 & 1 & 1 & 0 & 0 & 1 & 0 & 0 & -1 & -1 & 0 \\
 1 & -1 & 0 & 0 & 2 & -2 & -2 & 0 & 1 & 2 & 0 \\
 1 & -2 & -1 & 0 & 3 & -4 & -3 & 0 & 2 & 4 & 0 \\
 -1 & -1 & -1 & -1 & -1 & -1 & 1 & 1 & 0 & 0 & 0 \\
 -1 & -1 & -1 & -1 & -1 & -1 & 0 & 0 & 1 & 1 & 1 \\
    \end{array}
    \right)
\end{equation}

\begin{equation}
    \label{eqn_61_V}
    V =\left( \begin{array}{ccccccccccc}
    1 & 0 & 0 & 0 & 0 & 0 & 0 & -5 & 19 & -1 & -1 \\
 0 & 1 & 0 & 0 & 0 & 0 & 0 & -4 & 16 & -1 & -1 \\
 0 & 0 & 1 & 0 & 0 & 0 & 0 & -4 & 15 & -1 & -1 \\
 0 & 0 & 0 & 1 & 0 & 0 & 0 & -2 & 8 & -1 & -1 \\
 0 & 0 & 0 & 0 & 1 & 0 & 0 & -3 & 11 & -1 & -1 \\
 0 & 0 & 0 & 0 & 0 & 1 & 0 & -3 & 12 & -1 & -1 \\
 0 & 0 & 0 & 0 & 0 & 0 & 1 & 0 & 0 & 1 & 0 \\
 0 & 0 & 0 & 0 & 0 & 0 & 0 & 2 & -8 & 1 & 0 \\
 0 & 0 & 0 & 0 & 0 & 0 & 0 & 1 & -4 & 1 & 0 \\
 0 & 0 & 0 & 0 & 0 & 0 & 0 & 2 & -7 & 0 & 1 \\
 0 & 0 & 0 & 0 & 0 & 0 & 0 & 1 & -4 & 0 & 1 \\
    \end{array}
    \right)
\end{equation}

The {diagonal entry 6 in the} matrix $D$ (Eq.~(\ref{eqn_61_D})) shows there is one symmetry indicated stable topological index {per Eq.~(\ref{eqn_index})}.
Plugging these matrices into Eq.~(\ref{eqn_index}), we get Eq.~(\ref{eqn_index2}) in Sec.~\ref{sec:TRIM}. It is the mirror Chern number, determined by symmetries mod 6.

\subsection{\label{app_EBR_31} $p\bar{3}1'$}
The basis for band co-representations (columns) and the basis for coefficients of EBRs (rows) are
\begin{equation}
\label{eqn_basis_br_3}
    \left( E^{\Gamma}_{\frac12g}, E^{\Gamma}_{\frac12u}, E^{\Gamma}_{\frac32g}, E^{\Gamma}_{\frac32u}, E^{X}_{g}, E^{X}_{u}, E^{M}_{\frac12}, E^{M}_{\frac32} \right) ,
\end{equation}
{and}
\begin{equation}
\label{eqn_basis_ebr_3}
    \left( E^{1a}_{\frac12g}, E^{1a}_{\frac12u}, E^{1a}_{\frac32g}, E^{1a}_{\frac32u}, E^{2b}_{\frac12}, E^{2b}_{\frac32}, E^{3c}_{g}, E^{3c}_{u} \right) .
\end{equation}
In these bases, the EBR matrix is
\begin{equation}
    \label{eqn_31_A}
    A = \left(
\begin{array}{cccccccc}
 1 & 0 & 0 & 0 & 1 & 0 & 2 & 0 \\
 0 & 1 & 0 & 0 & 1 & 0 & 0 & 2 \\
 0 & 0 & 1 & 0 & 0 & 1 & 1 & 0 \\
 0 & 0 & 0 & 1 & 0 & 1 & 0 & 1 \\
 1 & 0 & 1 & 0 & 1 & 1 & 1 & 2 \\
 0 & 1 & 0 & 1 & 1 & 1 & 2 & 1 \\
 1 & 1 & 0 & 0 & 1 & 1 & 2 & 2 \\
 0 & 0 & 1 & 1 & 1 & 1 & 1 & 1 \\
\end{array}
\right)
\end{equation}
The Smith normal form matrices are
\begin{equation}
    \label{eqn_31_D}
    D = \left(
\begin{array}{cccccccc}
 1 & 0 & 0 & 0 & 0 & 0 & 0 & 0 \\
 0 & 1 & 0 & 0 & 0 & 0 & 0 & 0 \\
 0 & 0 & 1 & 0 & 0 & 0 & 0 & 0 \\
 0 & 0 & 0 & 1 & 0 & 0 & 0 & 0 \\
 0 & 0 & 0 & 0 & 1 & 0 & 0 & 0 \\
 0 & 0 & 0 & 0 & 0 & 2 & 0 & 0 \\
 0 & 0 & 0 & 0 & 0 & 0 & 0 & 0 \\
 0 & 0 & 0 & 0 & 0 & 0 & 0 & 0 \\
\end{array}
\right)
\end{equation}

\begin{equation}
    \label{eqn_31_U}
    U = \left(
\begin{array}{cccccccc}
 0 & 0 & 0 & 1 & 0 & -1 & 1 & 0 \\
 -1 & 0 & 0 & 0 & 0 & 0 & 1 & 0 \\
 0 & 0 & 1 & 0 & 0 & 0 & 0 & 0 \\
 0 & 0 & 0 & 1 & 0 & 0 & 0 & 0 \\
 1 & 1 & 0 & 0 & 0 & 0 & -1 & 0 \\
 0 & -1 & 0 & -1 & 0 & 1 & 0 & 0 \\
 -1 & -1 & -1 & -1 & 1 & 1 & 0 & 0 \\
 -1 & -1 & -1 & -1 & 0 & 0 & 1 & 1 \\
\end{array}
\right)
\end{equation}

\begin{equation}
    \label{eqn_31_V}
    V = \left(
\begin{array}{cccccccc}
 1 & 0 & 0 & 0 & 0 & 0 & -1 & -2 \\
 0 & 1 & 0 & 0 & 0 & 0 & -1 & -2 \\
 0 & 0 & 1 & 0 & 0 & -1 & -1 & -1 \\
 0 & 0 & 0 & 1 & 0 & 0 & -1 & -1 \\
 0 & 0 & 0 & 0 & 1 & 0 & 1 & 0 \\
 0 & 0 & 0 & 0 & 0 & 0 & 1 & 0 \\
 0 & 0 & 0 & 0 & 0 & 1 & 0 & 1 \\
 0 & 0 & 0 & 0 & 0 & 0 & 0 & 1 \\
\end{array}
\right)
\end{equation}

The {diagonal entry 2 in the} matrix $D$ (Eq.~(\ref{eqn_31_D})) shows there is one symmetry indicated stable topological index {per Eq.~(\ref{eqn_index})}.
Plugging these matrices into Eq.~(\ref{eqn_index}), we get Eq.~(\ref{eqn_index3}) in Sec.~\ref{sec:TRIM}. It is the mod 2 strong topological insulator index.

\section{\label{app_models}Bulk and surface band structures and parameter choices for tight-binding models}


In this section, we present more information about the two models that we presented in Sec.~\ref{sec:example}.\par
For both models, the parameters are chosen to satisfy $|m+6\tilde{M}_2|<|2\tilde{M}_1|$, so that there is one Dirac point along the high symmetry line $\Delta(\Gamma A)$ (and a pair of Dirac points inside the first BZ). 
The parameters for the first model as shown in Sec.~\ref{sec:example_NaBi} are $\tilde{M_2}=-1$, $m=4$, $\tilde{M_1}=-2.5$, $\tilde{A}=1$, $\tilde{B}=5$.
The parameters for the second model as shown in Sec.~\ref{sec:example_noFA} are: $\tilde{M_2}=-1$, $m=4$, $\tilde{M_1}=-2.5$, $\tilde{A}=5$, $\tilde{B_1}=1$, $\tilde{B_2}=1$. \par

\begin{figure}
    \centering
    \includegraphics[width=\linewidth]{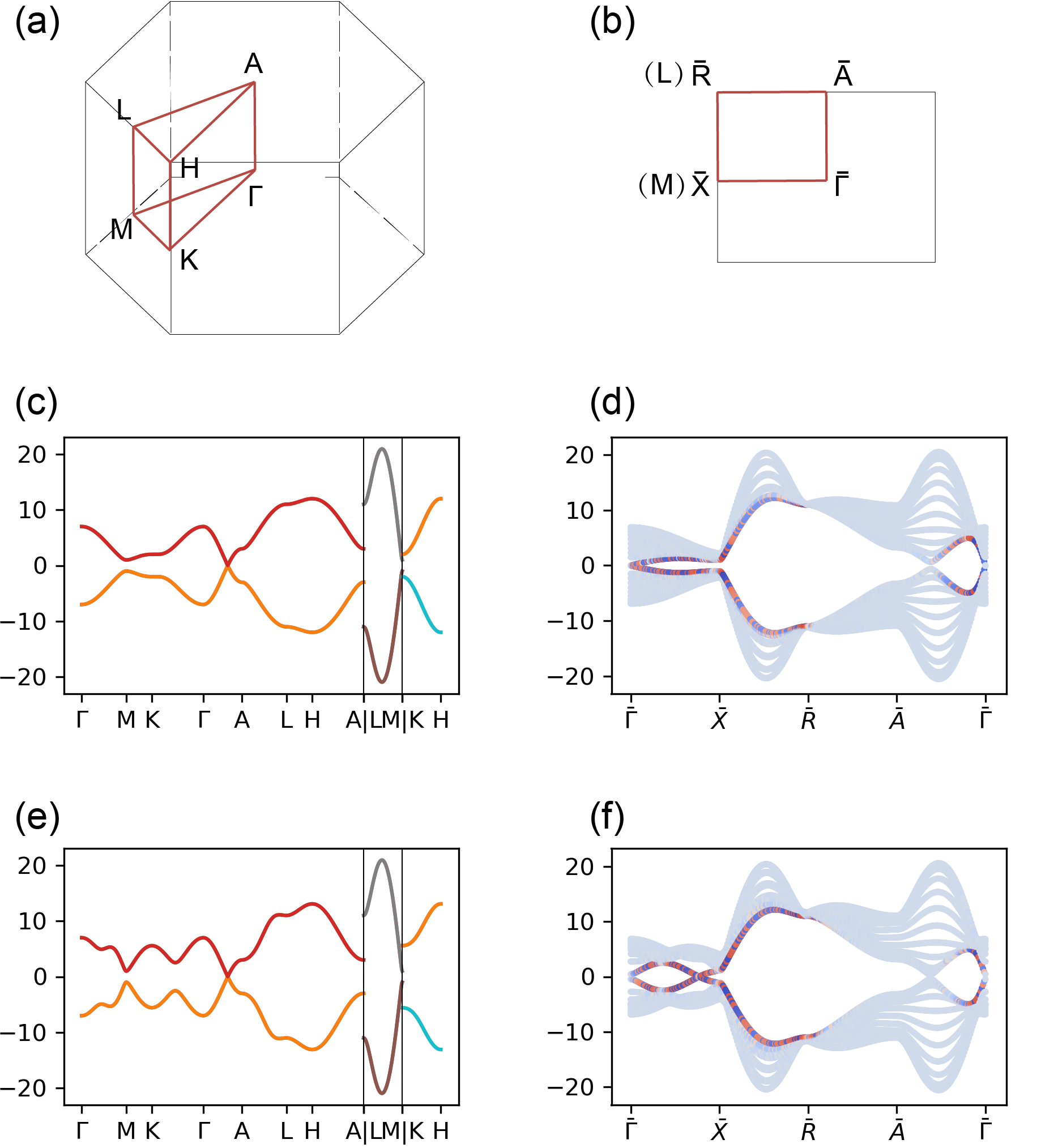}
    \caption{{Bulk and surface band structures for models of Dirac points in $P6/mmm$.} (a) The bulk BZ. (b) The surface BZ. This is a side surface with the normal $a_1=(1,0,0)$. 
    The $L$ and $M$ points of the bulk BZ are projected to $\bar{R}$ and $\bar{X}$ respectively. (c) Bulk spectrum of the first model Eq.(~\ref{eqn_model1_Ham}).
    (e) Bulk spectrum of the second model. (f) Surface spectrum of the second model Eq.~(\ref{eqn_model2_Ham}).}
    \label{fig_supp}
\end{figure}
\par
The two models both preserve space group $P6/mmm$. The unit cell of this space group is shown in Figure~\ref{fig_lattice}(a). The bulk BZ is shown in Figure~\ref{fig_supp}(a). The bulk spectrum for the two models are shown in Figures~\ref{fig_supp}(c) and (e) respectively.\par
We terminate the 3D lattice in the $a_1=(1,0,0)$ direction,
but keep the remaining directions infinite. 
The corresponding surface BZ is shown in Figure~\ref{fig_lattice}(b). The surface spectrum for the two models are shown in Figures~\ref{fig_supp}(d) and (f) respectively. These surface states 
{impact the rod band structure because they project}
into the rod spectrum when the crystal is terminated in the $a_1$ and $a_2$ directions as Figure~\ref{fig_lattice}(b) shows. The computation of bulk and surface spectra are implemented with Python package PythTB~\cite{pythtb}. \par

The high symmetry points (HSPs) that we are interested in are $\Gamma=(0,0,0)$, $A=(0,0,\pi)$, $K=(4\pi/3,0,0)$, $H=(4\pi/3,0,\pi)$, $M=(\pi,-\pi/\sqrt{3},0)$, and $L=(\pi,-\pi/\sqrt{3},\pi)$. The co-representations of the occupied bands are the labelled by the symmetry co-representations of the little co-groups at those momenta. 
We can use Eq.~(\ref{eqn_index2}) to determine the stable index, {which is the} mirror Chern number $C_m$, at the two TRIM planes. The high symmetry lines {that can host Dirac points} are $\Delta=\Gamma A$ and $P=KH$. The co-representations at the high symmetry lines can be determined by compatibility condition with the symmetry co-representations at the HSP. We can use Eq.~(\ref{eqn_eta2}) to determine the filling anomalies at the high symmetry lines. Finally, notice there is a Dirac point at one of the high symmetry lines for each model. The co-irreps of the crossing bands can be seen from the Tables~\ref{tab:BR1} and \ref{tab:BR2},
which show that the occupied bands have different co-reps at $\Gamma$ and $A$. Therefore, there must be a Dirac point at some momenta $\pm DP=(0,0,\pm k_{0})$ on the high symmetry line $\Delta$.
\par

The symmetry co-representations of the occupied bands at HSPs in the first model is shown in Table~\ref{tab:BR1}. 
From the symmetry indicators for the stable 2D topological index and the filling anomaly, we conclude that: the stable index at $k_z=0$ plane is $C_m=1 \mod 6$ ($\nu=1 \mod 2$) {per Eq.~(\ref{eqn_index2})}; the stable index at $k_z=\pi$ plane is $C_m=0 \mod 6$ {per Eq.~(\ref{eqn_index2})}; the filling anomaly between $\Gamma-DP$ is $\eta =2 \mod 6$ {per Eq.~(\ref{eqn_eta2})}; the filling anomaly between $DP-A$ is $\eta =0 \mod 6$ {per Eq.~(\ref{eqn_eta2})}.
\par
The symmetry co-representations of the occupied bands at HSPs in the second model is shown in Table~\ref{tab:BR2}.  From the symmetry indicators for the stable 2D topological index and the filling anomaly, we conclude that: the stable index at $k_z=0$ plane is $C_m=3 \mod 6$ ($\nu=1 \mod 2$) {per Eq.~(\ref{eqn_index2})}; the stable index at $k_z=\pi$ plane is $C_m=0 \mod 6$ {per Eq.~(\ref{eqn_index2})}; the filling anomaly between $\Gamma-DP$ is $\eta =0 \mod 6$ {per Eq.~(\ref{eqn_eta2})}; the filling anomaly between $DP-A$ is also $\eta =0 \mod 6$ {per Eq.~(\ref{eqn_eta2})}.
\par
The symmetry indicators we calculate here agree with the numerical computation of the rod spectrum of the two models as we plot in Figure~\ref{fig_spectrum}. The {change in filling anomalies and the change in stable topological indices} also agree with Table~\ref{tab_summary}.
\par

\begin{table}[]
    \centering
    \begin{tabular}{|c|cccccc|ccc|}
    \hline
         HSP & $E_{\frac12g}$ & $E_{\frac12u}$ & $E_{\frac32g}$ & $E_{\frac32u}$ & $E_{\frac52g}$ & $E_{\frac52u}$ &$E_{\frac12}$ & $E_{\frac32}$ &$E_{\frac52}$ \\
         \hline
         $\Gamma$ &&&1 &&& &&1&\\
         \hline
         $A$ &&1& &&& &1&&\\
         \hline
         \hline
         HSP & $E_{\frac12}$ & $E_{\frac32}$  & $E_{\frac52}$ & && &$E_{\frac12}$ & $E_{\frac32}$ &\\
         \hline
        $K$  &&&1 &&& &{1}&&\\
        \hline
         $H$  &&&1 &&& &{1}&&\\
         \hline
         \hline
         HSP & $E_{\frac12g}$ & $E_{\frac12u}$  &&&& &$E_\frac12$&&\\
         \hline
        $M$  &&1& &&& &1&&\\
        \hline
         $L$  &&1& &&& &1&&\\
         \hline
    \end{tabular}
    \caption{The symmetry co-representations of the occupied bands at HSPs in the first model. The last three columns are the co-reps at the nearby momentum $\mathbf{k}=\text{HSP}\pm \delta k_z$.} 
    \label{tab:BR1}
    \centering
    \begin{tabular}{|c|cccccc|ccc|}
    \hline
         HSP & $E_{\frac12g}$ & $E_{\frac12u}$ & $E_{\frac32g}$ & $E_{\frac32u}$ & $E_{\frac52g}$ & $E_{\frac52u}$ &$E_{\frac12}$ & $E_{\frac32}$ &$E_{\frac52}$ \\
         \hline
         $\Gamma$ &&& &&1& &&&1\\
         \hline
         $A$ &&1& &&& &1&&\\
         \hline
         \hline
         HSP & $E_{\frac12}$ & $E_{\frac32}$  & $E_{\frac52}$ & && &$E_{\frac12}$ & $E_{\frac32}$ &\\
         \hline
        $K$  &&&1 &&& &1&&\\
        \hline
         $H$  &&&1 &&& &1&&\\
         \hline
          \hline
         HSP & $E_{\frac12g}$ & $E_{\frac12u}$  &&&& &$E_\frac12$&&\\
         \hline
        $M$  &&1& &&& &1&&\\
        \hline
         $L$  &&1& &&& &1&&\\
         \hline
    \end{tabular}
    \caption{The symmetry co-representations of the occupied bands at HSPs in the second model.  The last three columns are the co-reps at the nearby momentum $\mathbf{k}=\text{HSP}\pm \delta k_z$.}
    \label{tab:BR2}
\end{table}
\par
\end{appendix}

\end{document}